\documentclass[10pt,preprint]{aastex}
\usepackage{float,amsmath,hyperref}
\usepackage{textcomp}
\usepackage{graphicx}

\newcommand{\teff}{$T_{\rm eff}$}
\newcommand{\tc}{$\rm T_{\mathrm{C}}$}
\newcommand{\ME}{$\rm M_{\oplus}$}
\newcommand{\MJ}{$\rm M_{\rm{J}}$}

\newcommand{\prv}{Pr0201}
\newcommand{\prw}{Pr0051}
\newcommand{\prx}{Pr0076}

\usepackage{ulem,color}

\begin{document}

\title{Detailed Abundances of Planet-Hosting Open Clusters.\
The Praesepe (Beehive) Cluster\altaffilmark{*}}
\author{
George Vejar\altaffilmark{1},
Simon C. Schuler\altaffilmark{2}, 
and Keivan G. Stassun\altaffilmark{1}% 
}
\affil{
%\altaffiltext{1}{Leibniz Institue for Astrophysics Potsdam (AIP); cmack@aip.de}
\altaffiltext{1}{Vanderbilt University, Department of Physics \& Astronomy, 6301 Stevenson Center Ln., Nashville, TN 37235, USA}
\altaffiltext{2}{University of Tampa, Department of Chemistry, Biochemistry, and Phyiscs, 401 W. Kennedy Blvd, Tampa, FL  33606, USA}
}

\altaffiltext{*}{The data presented herein were obtained at the W. M. Keck Observatory, which is operated as a scientific partnership among the California Institute of Technology, the University of California and the National Aeronautics and Space Administration. The Observatory was made possible by the generous financial support of the W. M. Keck Foundation.}

\begin{abstract}
It is not yet fully understood how planet formation affects the properties of host stars, 
in or out of a cluster; however, abundance trends can help us understand these processes.
We present a detailed chemical abundance analysis of six stars 
in Praesepe, a planet-hosting open cluster.
Pr0201 is known to host a close-in (period of 4.4 days) giant planet (mass of 0.54\MJ), while the other five cluster members in our sample
(Pr0133, Pr0081, Pr0208, Pr0051, and Pr0076) have no detected planets according to RV measurements.
Using high-resolution, high signal-to-noise echelle spectra
obtained with Keck/HIRES and a novel approach to equivalent width measurements (XSpect-EW), we derived abundances of up to 20 
elements spanning a range of condensation temperatures (\tc). We find a mean cluster metallicity of [Fe/H] = +0.21$\pm$0.02 dex, in agreement with most previous determinations.
We find most of our elements show a [X/Fe] scatter of $\sim$0.02-0.03 dex and conclude that our stellar sample is chemically homogeneous.
The \tc\ slope for the cluster mean abundances is consistent with zero and none of the stars in our sample exhibit individually a statistically significant \tc\ slope.
Using a planet engulfment model, we find that the planet-host, Pr0201, shows no evidence of significant enrichment in its refractory elements when compared to the cluster mean that would be consistent with a planetary accretion scenario.
\end{abstract}


\section{Introduction\label{s:intro}}
%{\bf Provide context for extending our work on planet-hosting wide binaries to planet-hosting open clusters.}

It has been over a decade now since the discoveries of \citet{2009ApJ...704L..66M} and \citet{2009A&A...508L..17R} suggesting that the formation of our solar system’s planets have imprinted a measurable trend on the elemental abundances of the Sun.
This trend, known as the condensation temperature (\tc) trend, can be influenced in various ways depending on the formation and evolution of the planetary system.
Planet formation may reduce abundances in refractory elements (\tc\ $>$ 900K), imparting a negative trend on abundances vs \tc\ since refractory-depleted material can still be accreted by the host star during the lifetime of the protoplanetary disk \citep{2017A&A...604L...4S}.
Planets in orbit can also be engulfed by their host star, which in turn could result in a positive slope since the star is accreting refractory-rich material into its outer layers \citep{2014ApJ...787...98M}.
The \tc\ slope can inform us on the amount of planetary material sequestered or accreted and help us determine if planet formation has occurred in the first place, in cases where planets have not been detected by other means.
Planetary signatures and \tc\ trends have been studied by numerous groups making use of wide binary systems with at least one known planet in order to take advantage of the assumption that they formed from the same molecular cloud and any differences in their abundances would be due to planet formation \citep{2011ApJ...740...76R,2014ApJ...787...98M,2014ApJ...790L..25T,2015A&A...583A.135B,2015ApJ...808...13R,2015A&A...582A..17S,2015ApJ...801L..10T,2016ApJ...818...54M,2017A&A...604L...4S,2019MNRAS.490.2448R}.
In this work, we apply similar methods to another chemically homogeneous stellar population: the Praesepe Open Cluster.

Open clusters are important laboratories for understanding a broad range of astrophysical phenomena. 
They have been used to study Galactic chemical evolution \citep{2020ApJ...888...28B, 2018AJ....155..138A}, the structure and evolution of the Galactic disk \citep{2015MNRAS.450.4301R, 2015MNRAS.446.3556M}, stellar physics \citep{2019ApJ...884..115D, 2009ApJ...701..837S}, and light element abundances \citep{2016ApJ...830...49B,2013A&A...552A.136F}, to name a few. 
The basis of all these studies is the assumption that open clusters are stellar conglomerates containing coeval stars that form out of a well-mixed molecular cloud. 
This implies that the stars in a given open cluster are the same age and have the same primordial compositions.
These properties allow for the systematic determination of their ages \citep{2020MNRAS.494.4713M,2016ApJ...831...11S}, distances \citep{2019MNRAS.487.2385M,2019A&A...626A..10G}, kinematic properties \citep{2020MNRAS.494.4713M,2015AJ....150...97G}, and detailed compositions \citep{2019ApJ...878...99L,2016MNRAS.463..696L}.
There may be more useful information available about the cluster environment than there would be about a wide binary system, such as more accurate age determinations.
A cluster can provide these advantages and many more stars to analyze in the context of exoplanets.

According to \citet{2013Natur.499...55M}, planets in stellar clusters may be just as likely as planets around field stars, but a cluster environment can also be hostile to the formation of planets and currently there are only tens of known planets in open clusters \citep{2019MNRAS.489.4311C}.
The main difference between field and cluster stars comes in the effect of the chaotic cluster environment on the formation process (protoplanetary disk) or already formed systems.
O/B stars emit high energy FUV photons that can photoevaporate nearby circumstellar disks, limiting planet formation timescales \citep{2013ApJ...774....9A,2018MNRAS.481..452H,2018MNRAS.478.2700W}.
Stellar fly-by's are frequent in the first 1-2 Myr after cluster formation which can lead to smaller systems ($<$ 5.5 AU in size) affecting 12-20$\%$ of stars in the lifetime of a cluster similar to Praesepe \citep{2018A&A...610A..33P}.
Flybys can also eject planets from a system with an efficiency of a few percent to $\sim$10$\%$ depending on semi-major axis of the planet, mass of the host star, and age of the cluster \citep{2019A&A...624A.110F}.
Gas expulsion from stellar winds of massive stars or supernova explosions can cause a cluster to become supervirial and boost ejection rates as the cluster reestablishes virial equilibrium \citep{2015MNRAS.453.2759Z}.
In low density environments (2k stars in 1pc virial radius), survival rates for systems containing multiple Jupiter-sized planets could be about $84\%$ and $90\%$ for Earth-only systems in the first 50 Myr \citep{2017MNRAS.470.4337C}.
Based off estimates of the specific-free floating planet production rate from \citet{2013ApJ...778L..42P}, Praesepe could have produced more than 1500 free floating planets. Aside from cluster environment, intra-system dynamics/evolution will also affect a fraction of surviving planetary systems such as: planetary migration \citep{1995Natur.378..355M,1996Natur.380..606L}, Kozai-Lidov effect \citep{2016ARA&A..54..441N}, and planet-planet scattering \citep{2012ApJ...758...39J} depending on the structure of the system.
The effectiveness of these mechanisms depends heavily on the size and structure of the cluster.

In this paper, we present the analysis of detailed abundance trends for six stars in Praesepe, a planet-hosting open cluster.
In Section~\ref{s:data}, we describe our sample, observations/data, spectral analysis including our novel approach to measuring absorption line EWs, and verify our methods.
In Section~\ref{s:results}, we compare our results to the literature and present the derived stellar abundances and observed trends. 
In Section~\ref{s:disc}, we discuss our results in the context of a simple model for how the accretion of Earth-like rocky planets would affect refractory elemental abundances as a function of \tc\ and atomic number.
Finally, we briefly summarize the main conclusions in Section~\ref{s:conc}.

\section{Data and Analysis\label{s:data}}

\subsection{Stellar Sample}
Stars in Praesepe are of special interest due to the discovery of planets around a number of its members \citep{2012ApJ...756L..33Q,2019MNRAS.489.4311C}.
Praesepe is home to $\sim$1000 stars, is relatively close by at 182 pc \citep{2018A&A...618A..93C}, and has an age of $\sim$600 Myr \citep{2011ASPC..448..841D}, making it a strong candidate for high resolution spectroscopy of main sequence sun-like stars.
It has the highest metallicity ([Fe/H] = +0.21 $\pm$ 0.01 dex) of any nearby open cluster according to \citet{2020AA...633A..38D}, which can increase the likelyhood of giant planet formation \citep{2010PASP..122..905J} if the metallicity correlation applies to stars in open clusters.

\begin{figure}[H]
\centering
\includegraphics[width=0.51\linewidth,trim=10 10 10 40,clip]{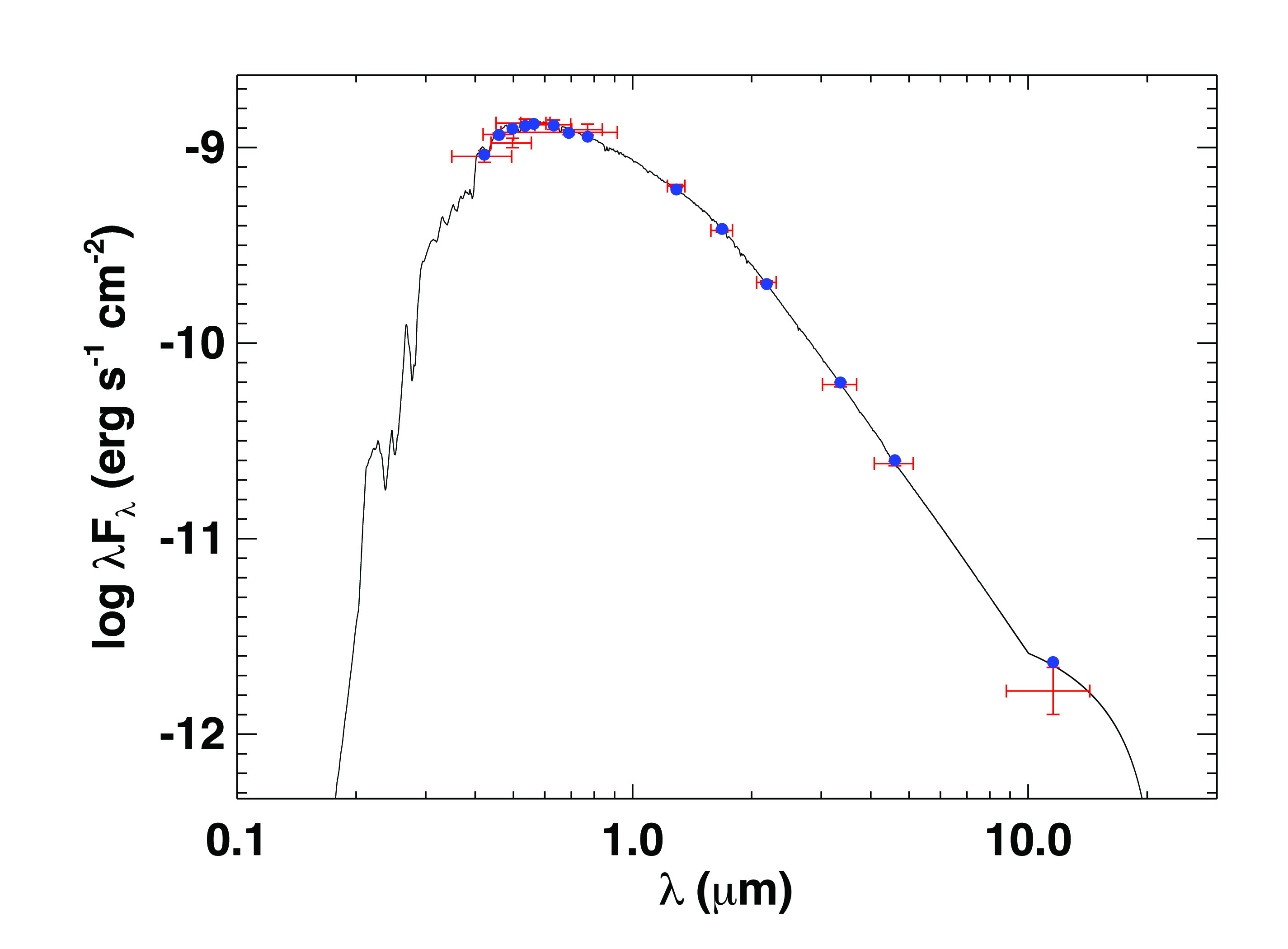}
\includegraphics[width=0.35\linewidth,trim=70 70 90 100,clip,angle=90]{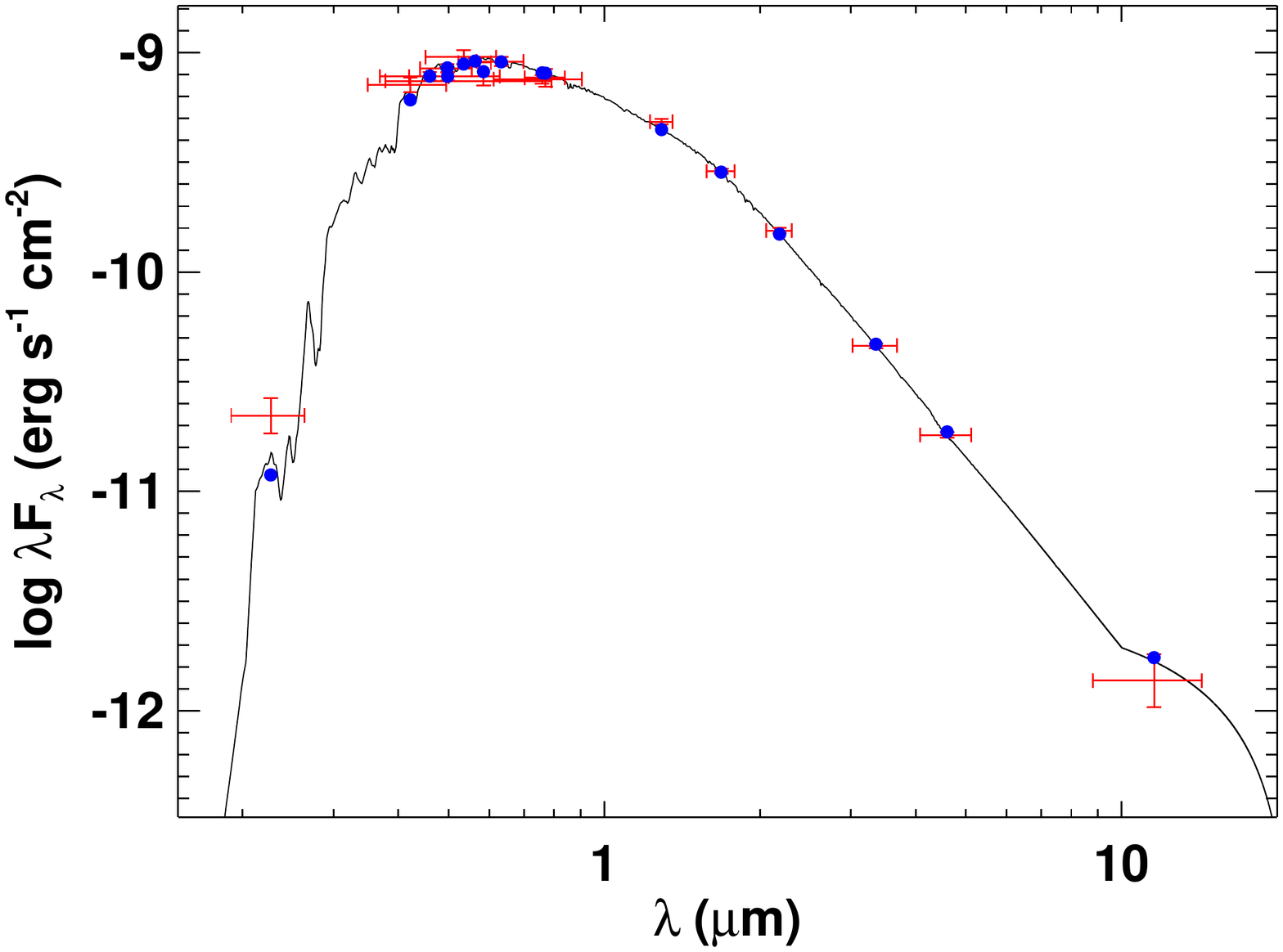}
\caption{Spectral Energy Distributions for two representative targets in our study sample. Red symbols represent the broadband fluxes drawn from GALEX \citep{GALEX:2003}, APASS \citep{APASS:2009}, 2MASS \citep{2MASS:2006}, and WISE \citep{WISE:2010}. Black curve is the best fitting Kurucz atmosphere model. Blue symbols are the model fluxes corresponding to each of the observed passbands. The integrated bolometric fluxes together with the {\it Gaia\/} DR2 parallax \citep{GaiaDR2:2018} yields the stellar radii. (Left) Pr0201: Reduced $\chi^{2}$ = 2.1, 
%adopting $\rm T_{\rm eff}$ = 6143 K from the table. The 
best fit extinction is $A_{V}$ = 0.072 $\pm$ 0.024, resulting in a bolometric flux at Earth of $F_{bol}$ = 1.77 $\pm$ 0.03 $\times 10^{-9}$ erg~s$^{-1}$~cm$^{-2}$, giving a radius of R = 1.166 $\pm$ 0.022 $R_{\odot}$. 
(Right) Pr0051:  Reduced $\chi^{2}$ = 1.2, 
%adopting $\rm T_{\rm eff} = 6017 \pm 27$~K and $\log g = 4.40 \pm 0.05$ and [Fe/H] = +0.16 $\pm$ 0.02 from the table. The 
best fit extinction is $A_{V}$ = 0.07 $\pm$ 0.03, resulting in a bolometric flux at Earth of $F_{\rm bol} = 1.241 \pm 0.029 \times 10^{-9}$ erg~s$^{-1}$~cm$^{-2}$, giving a radius of R = 1.086 $\pm$ 0.019 $R_{\odot}$. 
%The $\log g$ and radius together imply a mass of M = 1.08 $\pm$ 0.13 Msun. 
}
%\caption{\emph{Unweighted} linear fits to abundance vs.\ condensation temperature (\tc) for \hdab.}
\label{fig:SED}
\end{figure}

The six stars in our sample consist of a planet-host, Pr0201, and five non-hosts: Pr0133, Pr0081, Pr0076, Pr0051, and Pr0208.
The data used in this study are a combination of spectra acquired by our group in 2013 and high-quality Keck Observatory Archive (KOA) spectra taken at an earlier time, all of which are publicly available.
While other planet-hosts exist within Praesepe, we only acquired data for Pr0201.
The derived temperatures (see Sections 2.3 and 2.4) for our stars cover $\sim$500 K with spectral classes between G5 and F8 shown in Table \ref{tab:stellar_params}.
Half of our stars have temperatures within 100 K from the Sun and the other half are hotter with temperatures $\sim$6100 K.
Five out of the six stars have similar surface gravity estimates ranging from 4.34 to 4.44 dex and errors of about 0.10 dex.
Pr0133 is an exception having a much lower surface gravity of 4.18 dex.
The metallicity of our stars ranges between 0.16 and 0.26 with an average of $+0.21\pm 0.02$ dex.
In Figure \ref{fig:SED}, we show broadband SED fits \citep[using the methodology of][]{Stassun:2016} for two of our stars which are also consistent with our spectral analysis.

\begin{deluxetable}{lccccccc}
\tablecolumns{7}
\tablewidth{0pt}
\tabletypesize{\scriptsize}
%\tabletypesize{\tiny}
\tablecaption{Stellar Parameters\tablenotemark{a}\label{tab:stellar_params}}
\tablehead{
	\colhead{}&
	\colhead{Pr0201}&
	\colhead{Pr0133}&
	\colhead{Pr0208}&
	\colhead{Pr0081}&
	\colhead{Pr0051}&
	\colhead{Pr0076 (2003)}&
	\colhead{Pr0076 (2013)}
	}
\startdata
\teff (K)           & $6168\pm35$   & $6067\pm60$   & $5869\pm46$   & $5731\pm42$ 	& $6017\pm27$   & $5789\pm72$     & $5748\pm24$\\
$\log g$ (cgs)      & $4.34\pm0.10$  & $4.18\pm0.12$ & $4.37\pm0.13$ & $4.44\pm0.11$ & $4.40\pm0.07$ & $4.48\pm0.12$   & $4.44\pm0.07$\\
$[$Fe/H$]$ & $0.23\pm0.05$ & $0.19\pm0.06$ & $0.26\pm0.07$ & $0.18\pm0.06$ & $0.16\pm0.03$ & $0.25\pm0.06$ & $0.22\pm0.05$\\
$\xi$ (km s$^{-1}$) & $1.52\pm0.06$ & $1.74\pm0.11$ & $1.53\pm0.07$ & $1.41\pm0.06$ & $1.54\pm0.05$ & $1.33\pm0.04$   & $1.35\pm0.03$\\
\enddata
\tablenotetext{a}{Adopted solar parameters: \teff\ $=5777$ K, $\log g=4.44$, and $\xi=1.38$ km s$^{-1}$.}
\end{deluxetable}

\subsection{Data Acquisition}
Three of the six stars (\prv, \prw, and \prx) were observed on UT 2013 December 9 with the HIRES echelle spectrograph \citep{1994SPIE.2198..362V} in the $\rm R=\lambda/\Delta\lambda=72,000$ mode on the 10-m Keck I telescope. We used the kv418 filter combined with the B2 slit setting ($0\farcs574 \times 7\arcsec$) and $2 \times 1$ binning; the spectra cover a wavelength range of 4600--9000\AA. One exposure was taken for each star, with an integration time of 1200 s for \prv\ and 2100~s for \prw\ and \prx\ individually. The signal-to-noise ratio (S/N) in the continuum near 6700\AA\ for \prv\ and \prw\ is $\sim$300, and for \prx, it is $\sim$250.  

For the remaining three stars, Pr0133, Pr0208, and Pr0081, we obtained raw data files from the KOA.
These spectra, which we will collectively refer to as archive spectra, were taken in UT 2003 in January and February belonging to program ID H39aH and H47aH; the data are fully described in \citet{2013ApJ...775...58B}.
We also obtained an archive spectrum for Pr0076 from the same program to verify that our analysis produces consistent results between these different sets of spectra.
In our final results we report the stellar parameters and abundances for Pr0076 from both data sets and adopt the results from data taken in 2013.
The archive spectra covered a smaller wavelength range (5650--8090\AA) than our new spectra and were obtained in the $\rm R=48,000$ mode.
One exposure was taken for each star with an integration time of 840~s for Pr0133, 1200~s for Pr0208, 1500~s for Pr0076, and 1500~s for Pr0081.
The S/N in the continuum near 6700\AA\ for these spectra is $\sim$220. 

\begin{figure}[!ht]
\centering
\includegraphics[scale=0.67,trim=10 15 0 0,clip]{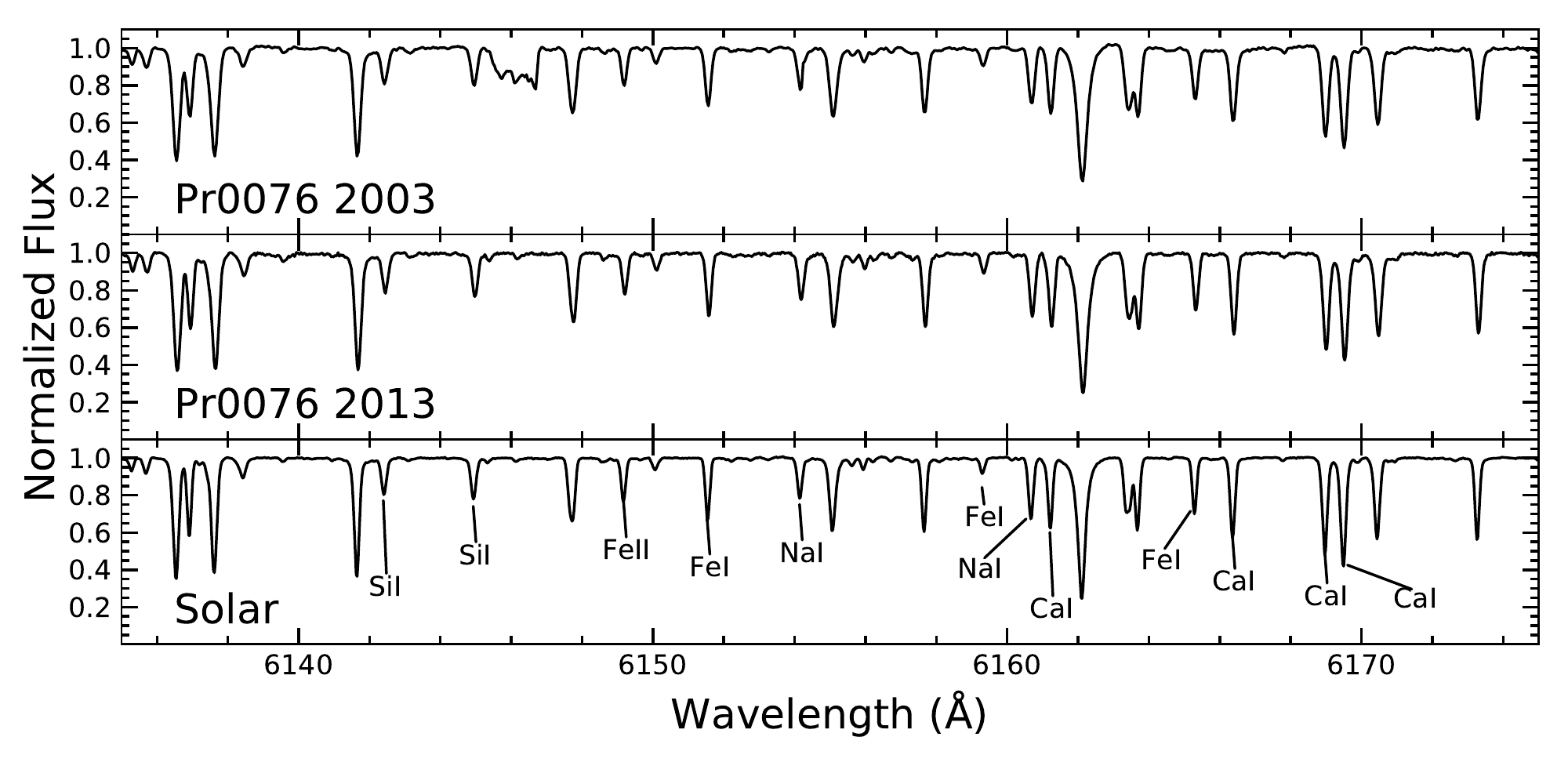}
\caption{Sample Keck/HIRES spectra for Pr0076 and the sun, spanning a wavelength range of $\sim6135-6175~$\AA. Top panel represents the archive data, middle panel represents data taken in 2013, and bottom panel shows the solar spectrum used. Marked absorption lines are those measured within this range.}
\label{fig:sample_spect}
\end{figure}

All of the data were reduced consistently using the MAKEE
%\footnote{\href{http://www.astro.caltech.edu/~tb/makee/}{http://www.astro.caltech.edu/$\sim$tb/makee/}} 
data reduction software. 
We required a solar spectrum to derive the solar abundances used to determine the abundances of our target stars relative to the Sun. For this purpose, we used the high-quality Keck/HIRES solar spectrum ($\mathrm{S/N} \sim 800$ near 6700\AA) from \citet{2015ApJ...815....5S} obtained in 2010 June in the R=50,000 mode over the wavelength range 3750--8170\AA.
A sample region of the spectra used is shown in Figure~\ref{fig:sample_spect} for Pr0076 from both data sets in comparison to the solar spectrum.

\subsection{Abundance and Stellar Parameter Determination}
We have derived chemical abundances relative to solar ([X/H]) for up to 20 elements in each of our stars. 
For our analysis, the adopted solar parameters were \teff\ $=5777$ K, $\log g=4.44$, and $\xi=1.38$ km s$^{-1}$.
%Derived stellar parameters are shown above in Table \ref{tab:stellar_params} and abundances are presented in Table \ref{tab:params}.
A sample of the adopted line lists, equivalent widths (EWs), and $\log(N)$ line-by-line abundances for each element are given in Table \ref{tab:linelist}. 
We derived abundances from measurements of EWs of atomic absorption lines in combination with MOOG, an LTE spectral analysis package \citep[version 2014]{1973PhDT.......180S} and ATLAS9 stellar atmosphere models \citep{1993KurCD..13.....K}. 
For the archive spectra, we used an abbreviated line list excluding lines not found in those spectra. 
To determine the stellar parameters of each star, we require the excitation and ionization balance of Fe I and Fe II lines. 
Atomic excitation energies ($\chi$) and transition probabilities ($\log gf$) were taken from \citet{2014ApJ...787...98M}. 
For the odd-Z elements Sc, V, Mn, and Co, hyper-fine structure (hfs) effects \citep{2000ApJ...537L..57P} are taken into account for strong lines through spectral synthesis incorporating hfs components; the resulting abundances are listed in Table \ref{tab:hfs_lines}.
The hfs components were obtained from \citet{2006ApJ...640..801J}, and line lists for regions encompassing each feature were taken from VALD \citep{1995A&AS..112..525P,1999A&AS..138..119K}.
Adopted Sc, V, Mn, and Co abundances were derived from the hfs analysis and lines with EWs where hfs effects are negligible. 
The general abundance and error analysis method is detailed in \citet{2011ApJ...737L..32S}, with new methods used for our current analysis detailed below.

%====================
%Linelist Table
%====================
\begin{deluxetable}{cccc|cc|cc|cc|cc|cc|cc|cc|}
\tablecolumns{16}
\tablewidth{0pt}
\rotate
\tabletypesize{\scriptsize}
\tablecaption{Lines Measured, Equivalent Widths, and Abundances\label{tab:linelist}}
\tablehead{
     \colhead{}&
     \colhead{$\lambda$}&
     \colhead{$\chi$}&
     \colhead{}&
     \multicolumn{2}{c}{Solar}&
     \multicolumn{2}{c}{Pr0201}&
     \multicolumn{2}{c}{Pr0133}&
     \multicolumn{2}{c}{Pr0208}&
     \multicolumn{2}{c}{Pr0081}&
     \multicolumn{2}{c}{Pr0051}&
     \multicolumn{2}{c}{Pr0076}\\
     %\cline{5-6} \cline{7-8} \cline{9-10} \cline{11-12}\\
     \colhead{Ion}&
     \colhead{(\AA)}&
     \colhead{(eV)}&
     \colhead{$\log \mathrm{gf}$}&
     \colhead{EW$_{\odot}$}&
     \colhead{$\log N_{\odot}$}&
     
     \colhead{EW}&
     \colhead{$\log N$}&
     \colhead{EW}&
     \colhead{$\log N$}&
     \colhead{EW}&
     \colhead{$\log N$}&
     \colhead{EW}&
     \colhead{$\log N$}&
     \colhead{EW}&
     \colhead{$\log N$}&
     \colhead{EW}&
     \colhead{$\log N$}
     }
\startdata

\ion{C}{1}  & 5052.167 & 7.685 & -1.304 & 31.58 & 8.417 & 48.52 & 8.481 & --    & --    & --    & --    & --    & --    & 44.66 & 8.507 & --    & --    \\
            & 5380.337 & 7.685 & -1.615 & 19.37 & 8.444 & 33.17 & 8.531 & --    & --    & --    & --    & --    & --    & 30.44 & 8.566 & 20.54 & 8.481 \\
            & 6587.61  & 8.537 & -1.021 & 12.2  & 8.358 & 28.46 & 8.565 & 20.64 & 8.387 & 20.58 & 8.567 & 14.05 & 8.454 & 19.24 & 8.43  & 12.22 & 8.358 \\
            & 7111.469 & 8.64  & -1.074 & 10.53 & 8.431 & 23.78 & 8.596 & 12.28 & 8.24  & 15.3  & 8.544 & 13.8  & 8.598 & --    & --    & 10.75 & 8.443 \\
            & 7113.179 & 8.647 & -0.762 & 22.81 & 8.563 & 40.39 & 8.642 & 30.7  & 8.482 & 26.88 & 8.581 & 26.74 & 8.694 & 30.7  & 8.561 & 24.98 & 8.632 \\

\ion{N}{1}  & 7468.313 & 10.336 & -0.189 & 4.18 & 8.131 & 9.48 & 8.195 & -- & -- & -- & -- & -- & -- & 7.64 & 8.21 & -- & -- \\

\ion{O}{1} & 6300.304 & 0.0 & -9.72 & 5.10  & 8.838 & 4.96  & 8.930 & --    & --    & --    & --    & --    & --    & --    & --    & 8.44  & 9.141\\
		   & 7775.39 & 9.15 & 0.001 & 45.45 & 8.806 & 84.57 & 9.078 & 84.38 & 9.139 & 61.74 & 9.007 & 58.58 & 9.097 & 67.51 & 8.951 & 51.79 & 8.96 \\
		   & 7774.17 & 9.15 & 0.223 & 64.78 & 8.925 & 105.5 & 9.139 & 95.89 & 9.082 & 77.69 & 9.039 & 64.0  & 8.968 & 86.72 & 9.013 & 64.3  & 8.96 \\
           & 7771.94 & 9.15 & 0.369 & 69.65 & 8.856 & 119.2 & 9.16  & 106.3 & 9.076 & 86.34 & 9.018 & 72.46 & 8.956 & 98.46 & 9.024 & 74.15 & 8.97 \\

\ion{Na}{1} & 5682.633 & 2.102 & -0.7  & 98.56 & 6.282 & 102.5 & 6.552 & --    & --    & 116.8 & 6.555 & --    & --	   & 99.82 & 6.419 & 119.2 & 6.524 \\
		    & 6154.226 & 2.102 & -1.56 & 36.15 & 6.267 & 35.74 & 6.449 & 35.3  & 6.359 & 47.67 & 6.49  & 45.52 & 6.391 & 37.5  & 6.403 & 49.84 & 6.466 \\	
			& 6160.747 & 2.104 & -1.26 & 54.69 & 6.255 & 49.66 & 6.374 & 51.33 & 6.317 & 64.53 & 6.43  & 61.46 & 6.321 & 54.25 & 6.362 & 67.11 & 6.411 \\	

\ion{Mg}{1} & 4730.029 & 4.346 & -2.523 & 66.13 & 7.802 & 60.68 & 7.927 & --    & --    & --    & --    & -- & -- & 67.72 & 7.942 & 81.46 & 8.009 \\
			& 5711.088 & 4.346 & -1.833 & 103.6 & 7.597 & 96.82 & 7.725 & --    & --    & 117.8 & 7.827 & -- & -- & 103.6 & 7.718 & 119.7 & 7.798 \\
    	    & 6965.409 & 5.753 & -1.51  & --    & --    & --    & --    & --    & --    & 25.88 & 7.34  & -- & -- & --    & --    & --    & --    \\
    	    & 6841.19  & 5.753 & -1.61  & 64.3  & 7.849 & 69.54 & 8.004 & 77.19 & 8.063 & 85.91 & 8.139 & -- & -- & 68.45 & 7.942 & 74.84 & 8.08  \\

\ion{Al}{1} & 6696.023 & 3.143 & -1.347 & 37.33 & 6.253 & --    & --    & --   & --    & 47.45 & 6.453 & 44.71 & 6.348 & 39.17 & 6.39  & 51.94 & 6.47 \\
            & 6698.673 & 3.143 & -1.647 & 21.1  & 6.222 & 18.94 & 6.339 & 20.1 & 6.292 & 27.89 & 6.413 & 27.36 & 6.34  & 22.68 & 6.366 & 29.37 & 6.39 \\
                
\ion{Si}{1} & 5701.104 & 4.93  & -1.581 & 37.85 & 7.087 & 39.86 & 7.236  & --    & --    & --    & --    & --    & --    & 41.87 & 7.216 & 47.47 & 7.266 \\
		    & 5690.425 & 4.93  & -1.769 & 50.48 & 7.485 & 54.33 & 7.66  & --    & --    & 59.0  & 7.641 & --    & --    & 51.69 & 7.563 & 57.96 & 7.621 \\   
		    & 5708.4   & 4.954 & -1.034 & 75.82 & 7.14  & 84.46 & 7.388 & --    & --    & 90.55 & 7.375 & --    & --    & 80.76 & 7.268 & 91.02 & 7.378 \\   
		    & 5772.149 & 5.082 & -1.358 & 50.81 & 7.213 & 58.97 & 7.445 & 59.77 & 7.384 & 66.03 & 7.466 & 62.82 & 7.402 & 56.25 & 7.35  & 63.21 & 7.42  \\ 
		    & 7405.772 & 5.614 & -0.313 & 89.3  & 7.108 & 100.7 & 7.352 & 102.1 & 7.332 & 104.6 & 7.326 & 106.4 & 7.324 & 100.7 & 7.289 & 106.3 & 7.34  \\ 
		    & 6125.021 & 5.614 & -1.464 & 32.55 & 7.48  & 39.8  & 7.688 & 39.12 & 7.62  & 41.65 & 7.658 & 39.71 & 7.614 & 38.83 & 7.631 & 41.96 & 7.657 \\ 
		    & 6244.466 & 5.616 & -1.093 & 44.9  & 7.316 & 54.31 & 7.54  & 53.57 & 7.477 & 58.57 & 7.539 & 55.23 & 7.479 & 53.51 & 7.484 & 56.47 & 7.505 \\ 
		    & 6243.815 & 5.616 & -1.242 & 44.62 & 7.46  & 53.32 & 7.675 & 55.52 & 7.655 & 57.63 & 7.675 & 53.49 & 7.604 & 48.09 & 7.555 & 62.3  & 7.734 \\ 
		    & 6145.016 & 5.616 & -1.31  & 39.69 & 7.45  & 46.7  & 7.644 & 42.13 & 7.518 & 53.6  & 7.685 & 43.08 & 7.516 & 44.76 & 7.572 & 48.16 & 7.6   \\ 
		    & 6142.483 & 5.619 & -1.295 & 33.85 & 7.339 & 39.73 & 7.522 & 35.73 & 7.397 & 41.73 & 7.495 & 39.28 & 7.443 & 38.38 & 7.459 & 43.42 & 7.516 \\ 
		    & 6848.58  & 5.863 & -1.524 & 17.25 & 7.401 & 21.13 & 7.572 & 19.82 & 7.485 & 23.61 & 7.584 & 23.82 & 7.582 & 19.27 & 7.491 & 24.55 & 7.602 \\ 
		    & 6414.98  & 5.871 & -1.035 & 46.89 & 7.484 & 55.2  & 7.671 & --    & --    & --    & --    & --    & --    & 54.52 & 7.622 & 59.12 & 7.665 \\
		    & 7003.569 & 5.964 & -0.937 & 57.28 & 7.606 & 61.99 & 7.74  & 68.65 & 7.794 & 77.06 & 7.873 & 66.17 & 7.734 & 59.56 & 7.669 & 62.59 & 7.697 \\
		    & 6741.628 & 5.984 & -1.428 & 15.06 & 7.344 & 20.56 & 7.56  & 19.67 & 7.49  & 22.05 & 7.553 & 19.79 & 7.491 & 18.54 & 7.477 & 20.47 & 7.511 \\

\ion{S}{1}  & 4694.113 & 6.525 & -1.77  & 11.0  & 7.311 & 13.81 & 7.27  & --    & --    & --    & --    & -- & -- & 16.24 & 7.419 & 13.26 & 7.423 \\
            & 4695.443 & 6.525 & -1.92  & 6.13  & 7.167 & 11.15 & 7.307 & --    & --    & --    & --    & -- & -- & 9.21  & 7.267 & 9.29  & 7.382 \\
            & 6757.171 & 7.87  & -0.31  & 14.14 & 7.197 & 34.72 & 7.514 & 28.69 & 7.393 & 23.76 & 7.438 & -- & -- & 25.91 & 7.402 & 19.24 & 7.394 \\

\ion{K}{1}  & 7698.98  & 0.0   & -0.17  & 155.6 & 5.278 & 156.3 & 5.617 & 164.0 & 5.611 & -- & -- & 171.8 & 5.397 & 164.8 & 5.55 & 174.6 & 5.457 \\

\ion{Ca}{1} & 6572.779 & 0.0   & -4.24  & 33.23 & 6.252 & 22.26 & 6.404 & 26.57 & 6.319 & 45.12 & 6.535 & 47.48 & 6.438 & 28.67 & 6.397 & 47.77 & 6.467 \\
            & 6166.439 & 2.521 & -1.142 & 70.02 & 6.281 & 68.18 & 6.487 & 69.96 & 6.409 & 83.04 & 6.522 & 82.72 & 6.441 & 73.04 & 6.458 & 84.55 & 6.496 \\     
            & 6169.042 & 2.523 & -0.797 & 91.23 & 6.267 & 90.34 & 6.503 & 94.6  & 6.472 & 105.5 & 6.519 & 108.7 & 6.481 & 95.54 & 6.465 & 111.9 & 6.556 \\     
            & 6161.297 & 2.523 & -1.266 & 62.73 & 6.291 & 55.6  & 6.408 & 57.2  & 6.327 & 72.38 & 6.481 & 70.25 & 6.372 & 62.23 & 6.414 & 77.07 & 6.505 \\     
            & 6169.563 & 2.526 & -0.478 & 109.1 & 6.203 & 105.7 & 6.423 & 114.8 & 6.471 & 117.9 & 6.375 & 128.9 & 6.419 & 110.6 & 6.367 & 130.2 & 6.467 \\     
            & 7326.145 & 2.933 & -0.208 & 109.5 & 6.217 & 109.3 & 6.465 & --    & --    & --    & --    & --    & --    & 113.9 & 6.405 & 132.1 & 6.48  \\
            & 5867.562 & 2.933 & -1.57  & 23.79 & 6.288 & 21.84 & 6.448 & 23.51 & 6.396 & 32.43 & 6.517 & 32.54 & 6.446 & 24.85 & 6.439 & 36.48 & 6.533 \\   

\ion{Sc}{2} & 6604.601 &  1.357 & -1.309 & 36.77 & 3.217 & 41.01 & 3.319 & 39.38 &  3.124 & 43.33 & 3.377 & 41.49 & 3.362 & 40.08 & 3.295 & 43.39 &  3.415 \\
			& 6320.851 &  1.5   & -1.819 & 9.12  & 3.102 & 11.35 & 3.24  & --    &  --    & 11.43 & 3.261 & --    & --    & 9.38  & 3.144 & --    &  --    \\
			& 6245.637 &  1.507 & -1.03  & 34.46 & 3.058 & 38.73 & 3.158 & 36.45 &  2.952 & 39.43 & 3.191 & 35.59 & 3.135 & 35.55 & 3.094 & 38.06 &  3.2   \\

\ion{Ti}{1} & 5039.957 & 0.021 & -1.13   & 72.88 & 4.688 & 63.51 & 4.915 & --    & --    & --    & --    & --    & --    & 71.56 & 4.888 & 87.26 & 4.939 \\
			& 5210.385 & 0.048 & -0.884  & 87.96 & 4.747 & 77.02 & 4.925 & --    & --    & --    & --    & --    & --    & 84.93 & 4.898 & 102.0 & 4.989 \\
			& 5064.653 & 0.048 & -0.991  & 88.13 & 4.882 & 79.82 & 5.108 & --    & --    & --    & --    & --    & --    & 86.6  & 5.061 & --    & --    \\
			& 5024.844 & 0.818 & -0.602  & 70.96 & 4.924 & 65.67 & 5.181 & --    & --    & --    & --    & --    & --    & 68.36 & 5.074 & 84.73 & 5.167 \\
			& 5022.868 & 0.826 & -0.434  & 69.99 & 4.745 & 61.11 & 4.938 & --    & --    & --    & --    & --    & --    & 69.0  & 4.926 & 83.53 & 4.983 \\
			& 5739.469 & 2.249 & -0.6    & 8.24  & 4.825 & --    & --    & --    & --    & 11.67 & 5.061 & --    & --    & --    & --    & 12.64 & 5.01  \\
			& 5866.451 & 1.067 & -0.84   & 47.33 & 4.901 & 37.04 & 5.078 & 40.2  & 4.959 & 59.03 & 5.159 & 63.42 & 5.118 & 42.58 & 5.033 & 61.57 & 5.115 \\
			& 6261.098 & 1.43  & -0.479  & 48.81 & 4.903 & 40.65 & 5.097 & 42.08 & 4.956 & 58.6  & 5.122 & 59.96 & 5.034 & 44.9  & 5.035 & 62.91 & 5.112 \\
			& 6258.102 & 1.443 & -0.355  & 51.46 & 4.839 & 42.9  & 5.025 & 43.19 & 4.864 & 60.88 & 5.048 & 64.88 & 5.007 & 47.35 & 4.966 & 64.86 & 5.036 \\
			& 6091.171 & 2.267 & -0.423  & 12.84 & 4.864 & 12.15 & 5.127 & --    & --    & --    & --    & --    & --    & 12.33 & 5.024 & 23.92 & 5.172 \\             
			& 6098.658 & 3.062 & -0.01   & 5.72  & 4.835 & --    & --    & --    & --    & 8.8   & 5.093 & 10.24 & 5.079 & --    & --    & 11.15 & 5.132 \\
			& 7138.906 & 1.443 & -1.59   & 6.34  & 4.825 & --    & --    & --    & --    & --    & --    & --    & --    & --    & --    & 11.07 & 5.06  \\ 

\ion{Ti}{2} & 5154.068 & 1.566 & -1.75   & 73.94 & 5.045 & 86.31 & 5.273 & --    & --    & --    & --    & --    & --    & 83.21 & 5.197 & 80.56 & 5.253 \\ 
			& 5381.021 & 1.566 & -1.92   & 60.55 & 4.933 & 72.6  & 5.15  & --    & --    & --    & --    & --    & --    & 68.14 & 5.059 & 69.83 & 5.186 \\ 
			& 5336.786 & 1.582 & -1.59   & 71.81 & 4.841 & 83.84 & 5.057 & --    & --    & --    & --    & --    & --    & 79.92 & 4.969 & 78.87 & 5.055 \\ 
			& 4779.985 & 2.048 & -1.26   & 63.97 & 4.859 & 75.92 & 5.066 & --    & --    & --    & --    & --    & --    & 72.79 & 4.999 & 67.48 & 5.002 \\ 

\ion{V}{1}  & 6285.15  & 0.275 & -1.51   & 10.22 & 3.844 & 6.71  & 4.065 & --    & --    & 14.86 & 4.125 & 15.42 & 3.992 & 7.83  & 3.978 & 17.47 & 4.077 \\
            & 6251.827 & 0.287 & -1.34   & 13.46 & 3.824 & 8.9   & 4.041 & --    & --    & 17.94 & 4.065 & 22.2  & 4.029 & 11.25 & 3.995 & 24.85 & 4.114 \\
            & 6224.529 & 0.287 & -2.01   & 4.9   & 4.018 & --    & --    & --    & --    & --    & --    & --    & --    & --    & --    & 9.64  & 4.298 \\
            & 6243.105 & 0.301 & -0.98   & 28.1  & 3.879 & 18.71 & 4.067 & 22.53 & 3.97  & 35.54 & 4.112 & 43.06 & 4.102 & 23.92 & 4.041 & 43.85 & 4.141 \\
            & 6111.645 & 1.043 & -0.715  & 10.09 & 3.822 & 10.95 & 4.232 & 11.57 & 4.086 & 16.87 & 4.164 & 17.17 & 4.041 & 10.48 & 4.072 & 20.07 & 4.143 \\
            & 5737.059 & 1.064 & -0.74   & 11.14 & 3.938 & 7.52  & 4.119 & --    & --    & 13.46 & 4.118 & --    & --    & 10.11 & 4.121 & 15.13 & 4.064 \\
            & 5727.048 & 1.081 & -0.012  & 38.13 & 3.93  & 27.15 & 4.071 & --    & --    & 47.27 & 4.16  & --    & --    & 34.47 & 4.079 & 56.29 & 4.224 \\
            & 6081.441 & 1.051 & -0.579  & 14.95 & 3.892 & --    & --    & --    & --    & --    & --    & --    & --    & 12.94 & 4.048 & 27.06 & 4.187 \\ 
            & 6090.214 & 1.081 & -0.062  & 32.8  & 3.851 & 27.53 & 4.106 & --    & --    & --    & --    & --    & --    & 29.91 & 4.015 & 47.35 & 4.088 \\

\ion{Cr}{1} & 6330.091 & 0.941 & -2.92   & 27.03 & 5.631 & 19.96 & 5.835 & 22.6  & 5.722 & 34.39 & 5.859 & 38.79 & 5.81  & 24.57 & 5.804 & 40.18 & 5.858 \\
			& 7400.249 & 2.9   & -0.111  & 75.93 & 5.569 & 73.08 & 5.775 & 76.58 & 5.708 & 94.81 & 5.896 & 96.07 & 5.838 & 80.61 & 5.778 & 93.11 & 5.826 \\
			& 5787.918 & 3.322 & -0.083  & 45.16 & 5.51  & 45.24 & 5.744 & --    & --    & --    & --    & --    & --    & 50.71 & 5.739 & 62.14 & 5.789 \\  
			& 5783.85  & 3.322 & -0.295  & 44.18 & 5.705 & 40.72 & 5.875 & 45.2  & 5.846 & 58.87 & 5.997 & 57.73 & 5.904 & 47.73 & 5.901 & 59.62 & 5.959 \\
			& 5783.063 & 3.323 & -0.5    & 30.73 & 5.656 & 29.43 & 5.859 & 29.03 & 5.742 & --    & --    & 44.19 & 5.878 & 32.97 & 5.838 & 42.79 & 5.871 \\ 
			& 5702.306 & 3.449 & -0.667  & 23.0  & 5.77  & 21.88 & 5.966 & --    & --    & --    & --    & --    & --    & 26.3  & 5.983 & 36.76 & 6.049 \\ 
			& 6978.397 & 3.464 & 0.142   & 58.64 & 5.594 & 59.81 & 5.839 & 58.49 & 5.712 & 77.04 & 5.923 & 75.88 & 5.834 & 61.86 & 5.771 & 76.98 & 5.878 \\
			& 6979.795 & 3.464 & -0.41   & 35.36 & 5.74  & 33.2  & 5.921 & 32.11 & 5.794 & 43.8  & 5.935 & 47.11 & 5.919 & 37.75 & 5.916 & 50.97 & 6.002 \\

\ion{Mn}{1} & 5432.546 & 0.0   & -3.795  & 50.91 & 5.465 & 30.39 & 5.538  & --   & --    & --    & --    & --    & --    & 39.95 & 5.54  & 69.84 & 5.771 \\
			& 5399.499 & 3.853 & -0.287  & 37.61 & 5.518 & 37.5  & 5.739  & --   & --    & --    & --    & --    & --    & 39.15 & 5.677 & 52.82 & 5.785 \\
			
\ion{Fe}{1} & 4779.439 & 3.42  & -2.02   & 39.79 & 7.237 & 38.25 & 7.464  & --    & --    & --    & --     & --	  & --	  & 41.27  & 7.265 & 49.46  &  7.415    \\       
            & 4788.757 & 3.24  & -1.76   & 65.33 & 7.29  & 65.69 & 7.558  & --    & --    & --    & --     & --	  & --	  & 67.74  & 7.337 & 75.06  &  7.480    \\       
            & 5054.643 & 3.64  & -1.92   & 39.03 & 7.315 & 38.24 & 7.544  & --    & --    & --    & --     & --	  & --	  & 38.66  & 7.308 & 49.90  &  7.513    \\       
            & 5322.041 & 2.28  & -2.8    & 61.14 & 7.267 & 56.99 & 7.503  & --    & --    & --    & --     & --	  & --	  & 60.04  & 7.246 & --    & --    \\          
            & 5379.574 & 3.69  & -1.51   & 61.2  & 7.344 & 63.69 & 7.629  & --    & --    & --    & --     & --	  & --	  & 62.16  & 7.361 & 71.26  &  7.526 \\       
            & 5522.447 & 4.21  & -1.55   & 43.66 & 7.55  & 44.22 & 7.775  & --    & --    & --    & --     & --	  & --	  & 45.5   & 7.583 & 55.06  &  7.752 \\       
            & 5543.936 & 4.22  & -1.14   & 61.98 & 7.469 & 62.55 & 7.697  & --    & --    & --    & --     & --	  & --	  & 65.73  & 7.534 & 74.13  &  7.683 \\       
            & 5546.5   & 4.37  & -1.31   & 49.71 & 7.563 & 54.14 & 7.849  & --    & --    & --    & --     & --	  & --	  & 53.6   & 7.631 & 65.04  &  7.830 \\       
            & 5546.991 & 4.22  & -1.91   & 25.65 & 7.55  & 25.80 & 7.767  & --    & --    & --    & --     & --	  & --	  & 27.29  & 7.587 & 37.93  &  7.801 \\       
            & 5560.207 & 4.43  & -1.19   & 51.79 & 7.539 & 51.81 & 7.744  & --    & --    & --    & --     & --	  & --	  & 54.52  & 7.586 & 63.11  &  7.735 \\       
            & 5577.03  & 5.03  & -1.55   & 10.15 & 7.459 & 12.23 & 7.718  & --    & --    & --    & --     & --	  & --	  & 12.11  & 7.545 & 17.71  &  7.734 \\       
            & 5579.335 & 4.23  & -2.4    & 8.92  & 7.492 &  9.60 & 7.741  & --    & --    & --    & --     & --	  & --	  & 10.54  & 7.573 & 13.65  &  7.695 \\       
            & 5587.574 & 4.14  & -1.85   & 32.77 & 7.566 & 34.89 & 7.827  & --    & --    & --    & --     & --	  & --	  & 35.34  & 7.618 & 42.96  &  7.758 \\       
            & 5646.684 & 4.26  & -2.5    & 7.54  & 7.54  &  7.28 & 7.735  & --    & --    & --    & --     & --	  & --	  & 7.68   & 7.548 & 12.24  &  7.768 \\       
            & 5651.469 & 4.47  & -2.0    & 18.0  & 7.674 & 18.73 & 7.896  & --    & --    & --    & --     & --	  & --	  & 19.4   & 7.715 & 26.76  &  7.895 \\       
            & 5652.318 & 4.26  & -1.95   & 26.79 & 7.652 & 27.59 & 7.882  & --    & --    & --    & --     & --	  & --	  & 28.11  & 7.682 & 36.45  &  7.848 \\       
            & 5661.346 & 4.28  & -1.74   & 22.53 & 7.357 & 23.17 & 7.584  & --    & --    & --    & --     & --	  & --	  & 24.83  & 7.414 & 32.05  &  7.567 \\       
            & 5667.518 & 4.18  & -1.58   & 50.25 & 7.66  & 53.20 & 7.929  & --    & --    & --    & --     & --	  & --	  & 53.97  & 7.726 & 64.38  &  7.908 \\       
            & 5677.684 & 4.1   & -2.7    & 6.7   & 7.53  &  7.39 & 7.797  & --	& --	& 10.95 & 7.817  & --	  & --	  & --    & --     & 10.98  &  7.759 \\        
            & 5679.023 & 4.65  & -0.92   & 58.87 & 7.584 & 60.58 & 7.813  & --	& --	& 71.40 & 7.827  & --	  & --	  & 61.95  & 7.635 & 72.62  &  7.816 \\       
            & 5680.24  & 4.19  & -2.58   & 9.92  & 7.683 &  9.75 & 7.891  & --	& --	& 14.18 & 7.911  & --	  & --	  & 9.05   & 7.638 & 15.66  &  7.906 \\       
            & 5731.762 & 4.26  & -1.3    & 57.03 & 7.576 & 56.69 & 7.783  & --	& --	& 66.22 & 7.768  & --	  & --	  & 61.82  & 7.659 & 71.12  &  7.821 \\       
            & 5732.275 & 4.99  & -1.56   & 13.43 & 7.567 & 15.11 & 7.799  & --	& --	& 17.84 & 7.753  & --	  & --	  & 16.41  & 7.67  & 19.35  &  7.753 \\       
            & 5741.846 & 4.26  & -1.85   & 31.14 & 7.643 & 34.40 & 7.921  & --	& --	& 40.82 & 7.874  & --	  & --	  & 34.98  & 7.721 & 43.17  &  7.871 \\       
            & 5752.032 & 4.55  & -1.18   & 54.08 & 7.667 & 54.90 & 7.882  & --	& --	& 68.86 & 7.950  & --	  & --	  & 56.98  & 7.717 & 65.96  &  7.870 \\       
            & 5775.081 & 4.22  & -1.3    & 58.13 & 7.556 & 59.23 & 7.790  & 61.45 & 7.730 & 72.73 & 7.840  & 71.63  & 7.764 & 61.02  & 7.606 & 71.72  &  7.793 \\            
            & 5778.45  & 2.59  & -3.48   & 21.27 & 7.42  & 18.63 & 7.658  & 21.06 & 7.633 & 26.92 & 7.626  & 30.49  & 7.602 & 21.1   & 7.415 & 32.68  &  7.668 \\     
            & 5809.218 & 3.88  & -1.84   & 49.01 & 7.61  & 47.99 & 7.822  & 52.62 & 7.807 & 62.60 & 7.885  & 63.09  & 7.832 & 50.32  & 7.634 & 61.91  &  7.836 \\     
            & 5934.655 & 3.93  & -1.17   & 74.38 & 7.431 & 75.30 & 7.679  & 76.95 & 7.584 & 89.73 & 7.722  & 86.91  & 7.614 & 76.91  & 7.475 & 89.22  &  7.687 \\     
            & 6078.999 & 4.65  & -1.12   & 45.07 & 7.535 & 50.94 & 7.831  & --    & --    & --    & --     & --	  & --	  & 49.91  & 7.62  & 56.95  &  7.740 \\       
            & 6085.259 & 2.76  & -3.1    & 42.75 & 7.646 & 38.93 & 7.869  & --    & --    & --    & --     & --	  & --	  & 40.59  & 7.605 & 55.38  &  7.860 \\       
            & 6098.245 & 4.56  & -1.88   & 15.83 & 7.553 & 18.10 & 7.819  & 19.53 & 7.803 & 25.52 & 7.863  & 23.83  & 7.762 & 18.08  & 7.623 & 25.01  &  7.800 \\       
            & 6151.617 & 2.18  & -3.3    & 49.27 & 7.384 & 43.93 & 7.616  & 47.49 & 7.559 & 57.73 & 7.584  & 58.39  & 7.499 & 48.06  & 7.362 & 61.57  &  7.585 \\     
            & 6159.368 & 4.61  & -1.97   & 12.49 & 7.568 & 14.80 & 7.846  & 16.39 & 7.842 & 23.85 & 7.957  & 17.56  & 7.727 & 14.42  & 7.641 & 19.69  &  7.800 \\     
            & 6165.36  & 4.14  & -1.47   & 43.68 & 7.373 & 42.54 & 7.569  & 46.05 & 7.547 & 54.50 & 7.601  & 53.80  & 7.528 & 47.24  & 7.437 & 55.32  &  7.576 \\     
            & 6173.336 & 2.22  & -2.88   & 67.93 & 7.337 & 65.82 & 7.612  & 72.32 & 7.575 & 80.69 & 7.596  & 83.04  & 7.558 & 68.7   & 7.351 & --	& -- \\       
            & 6187.987 & 3.94  & -1.72   & 46.71 & 7.487 & 44.33 & 7.671  & --    & --    & --    & --     & --	  & --	  & 48.54  & 7.52  & 61.21  &  7.739 \\       
            & 6220.776 & 3.88  & -2.46   & 19.45 & 7.588 & 14.92 & 7.681  & 14.36 & 7.593 & 23.50 & 7.747  & 23.80  & 7.680 & 19.8   & 7.598 & 26.27  &  7.753 \\       
            & 6226.73  & 3.88  & -2.22   & 28.6  & 7.573 & 26.42 & 7.757  & 32.41 & 7.807 & 35.94 & 7.770  & 40.57  & 7.787 & 29.98  & 7.603 & 40.97  &  7.813 \\     
            & 6229.228 & 2.85  & -2.81   & 37.66 & 7.34  & 36.74 & 7.613  & 38.93 & 7.554 & 49.51 & 7.607  & 47.29  & 7.482 & 39.5   & 7.375 & 51.30  &  7.577 \\     
            & 6240.645 & 2.22  & -3.23   & 49.7  & 7.353 & 43.88 & 7.574  & 47.22 & 7.515 & 61.27 & 7.603  & 57.48  & 7.444 & 46.97  & 7.304 & 61.60  &  7.546 \\     
            & 6293.924 & 4.84  & -1.72   & 13.4  & 7.567 & 16.08 & 7.841  & --    & --    & --    & --     & --	  & --	  & 17.19  & 7.697 & 21.86  &  7.823 \\       
            & 6297.793 & 2.22  & -2.74   & 72.68 & 7.277 & 69.31 & 7.528  & --    & --    & --    & --     & --     & --    & 74.83  & 7.317 & --    & --    \\          
            & 6322.685 & 2.59  & -2.43   & 75.11 & 7.374 & 73.58 & 7.637  & 75.98	& 7.535	& 93.21 & 7.725  & 88.04  & 7.561 & 78.5   & 7.436 & 93.00  &  7.685 \\     
            & 6380.743 & 4.19  & -1.38   & 52.09 & 7.473 & 52.38 & 7.692  & 54.71	& 7.640	& 62.72 & 7.688  & 61.46  & 7.609 & 54.09  & 7.508 & 63.76  &  7.673 \\     
            & 6392.538 & 2.28  & -4.03   & 16.51 & 7.491 & 11.36 & 7.634  & 15.95	& 7.710	& 21.32 & 7.703  & 22.94  & 7.636 & 13.95  & 7.404 & 23.94  &  7.679 \\     
            & 6597.557 & 4.79  & -1.07   & 43.46 & 7.567 & 43.98 & 7.761  & 48.28	& 7.763	& 56.30 & 7.820  & 53.77  & 7.726 & 46.52  & 7.621 & 56.49  &  7.790 \\     
            & 6608.024 & 2.28  & -4.03   & 16.74 & 7.486 & 16.50 & 7.810  & 16.67	& 7.721	& 21.75 & 7.702  & 24.26  & 7.655 & 15.1   & 7.433 & 26.47  &  7.726 \\     
            & 6609.11  & 2.56  & -2.69   & 65.5  & 7.411 & 63.50 & 7.671  & 67.39	& 7.602	& 80.11 & 7.696  & 81.71  & 7.648 & 64.31  & 7.389 & 80.18  &  7.657 \\     
            & 6627.54  & 4.55  & -1.68   & 27.39 & 7.63  & 29.37 & 7.869  & 32.19	& 7.860	& 37.53 & 7.876  & 35.11  & 7.771 & 28.94  & 7.664 & 38.38  &  7.848 \\     
            & 6653.85  & 4.15  & -2.52   & 10.09 & 7.552 & 11.73 & 7.845  & --	& --	& --    & --     & --	  & --	  & 11.33  & 7.609 & 15.14  &  7.748 \\      
            & 6703.567 & 2.76  & -3.16   & 37.17 & 7.566 & 31.54 & 7.753  & 35.81	& 7.738	& 47.06 & 7.799  & 48.00  & 7.726 & 36.14  & 7.546 & 50.64  &  7.795 \\     
            & 6710.316 & 1.49  & -4.88   & 15.91 & 7.513 & 10.90 & 7.706  & 13.84	& 7.719	& 23.01 & 7.798  & 25.97  & 7.740 & 13.74  & 7.439 & 24.77  &  7.734 \\     
            & 6713.745 & 4.8   & -1.6    & 20.94 & 7.63  & 23.80 & 7.884  & 22.24	& 7.788	& 29.59 & 7.872  & 30.95  & 7.846 & 23.89  & 7.705 & 31.10  &  7.861 \\     
            & 6716.222 & 4.58  & -1.92   & 15.11 & 7.564 & 15.93 & 7.785  & 15.08	& 7.699	& 21.11 & 7.787  & 20.08  & 7.699 & 17.36  & 7.637 & 22.43  &  7.772 \\     
            & 6725.353 & 4.1   & -2.3    & 17.17 & 7.549 & 16.88 & 7.761  & 17.32	& 7.708	& 24.44 & 7.794  & 26.13  & 7.764 & 18.2   & 7.58  & 25.17  &  7.756 \\     
            & 6726.666 & 4.61  & -1.13   & 45.83 & 7.493 & 48.91 & 7.739  & 51.63	& 7.706	& 59.60 & 7.759  & 58.57  & 7.688 & 50.25  & 7.569 & 60.41  &  7.740 \\     
            & 6733.151 & 4.64  & -1.58   & 26.82 & 7.599 & 28.65 & 7.830  & 29.50	& 7.784 & 35.31 & 7.814  & 35.81  & 7.767 & 29.43  & 7.656 & 38.31  &  7.829 \\     
            & 6739.52  & 1.56  & -4.79   & 11.32 & 7.321 &  9.01 & 7.589  & --	& --	& 18.43 & 7.656  & 16.99  & 7.487 & 10.47  & 7.283 & 19.18  &  7.572 \\     
            & 6745.09  & 4.58  & -2.16   & 7.8   & 7.477 &  9.17 & 7.749  & --	& --	& 14.37 & 7.822  & --	  & --	  & 8.13   & 7.497 & 13.17  &  7.726 \\      
            & 6745.957 & 4.08  & -2.77   & 6.66  & 7.531 &  8.09 & 7.847  & --	& --	& --    & --     & --	  & --	  & 7.96   & 7.616 & 11.74  &  7.796 \\       
            & 6750.15  & 2.42  & -2.62   & 73.4  & 7.332 & 69.88 & 7.570  & 72.88	& 7.476	& 85.02 & 7.559  & 87.06  & 7.518 & 73.7   & 7.337 & 88.39  &  7.581 \\     
            & 6752.716 & 4.64  & -1.3    & 35.72 & 7.505 & 37.19 & 7.724  & 42.02	& 7.743	& 51.91 & 7.830  & 49.08  & 7.727 & 39.67  & 7.58  & 49.57  &  7.753 \\       
            & 7114.549 & 2.69  & -4.01   & 7.57  & 7.455 &  6.74 & 7.709  & --	& --	& 11.19 & 7.715  & 14.01  & 7.725 & 9.48   & 7.562 & 11.81  &  7.656 \\     
            & 7284.835 & 4.14  & -1.75   & 41.49 & 7.561 & 39.72 & 7.744  & 46.28	& 7.777	& 55.41 & 7.838  & 52.31  & 7.726 & 43.28  & 7.593 & 53.79  &  7.771 \\     
\ion{Fe}{2} & 4620.521 & 2.828 & -3.315  & 53.41 & 7.435 & 67.36 & 7.641  & --	& --	& --    & --     & --	  & --	  & 64.51  & 7.68  & 57.82  &  7.608 \\        
            & 5197.577 & 3.23  & -2.348  & 80.04 & 7.387 &105.30 & 7.775  & --	& --	& --    & --     & --	  & --	  & 98.31  & 7.746 & 89.86  &  7.672 \\        
            & 5234.625 & 3.221 & -2.279  & 80.43 & 7.314 &105.70 & 7.701  & --	& --	& --    & --     & --	  & --	  & 98.12  & 7.662 & 89.54  &  7.585 \\        
            & 5264.812 & 3.23  & -3.133  & 39.51 & 7.308 & 57.43 & 7.571  & --	& --	& --    & --     & --	  & --	  & 53.91  & 7.617 & 50.27  &  7.616 \\        
            & 5414.073 & 3.221 & -3.645  & 25.69 & 7.486 & 40.63 & 7.730  & --	& --	& --    & --     & --	  & --	  & 36.77  & 7.748 & 31.12  &  7.687 \\        
            & 5425.257 & 3.199 & -3.39   & 39.8  & 7.536 & 52.66 & 7.695  & --	& --	& --    & --     & --	  & --	  & 49.66  & 7.746 & 46.86  &  7.762 \\        
            & 6084.111 & 3.199 & -3.881  & 20.4  & 7.542 & 31.34 & 7.728  & --	& --	& --    & --     & --	  & --	  & 27.42  & 7.73  & 25.58  &  7.752 \\        
            & 6113.322 & 3.221 & -4.23   & 12.35 & 7.628 & 20.74 & 7.831  & --	& --	& 17.90 & 7.841  & 13.98  & 7.756 & 16.98  & 7.8   & 15.65  &  7.820 \\    
            & 6149.258 & 3.889 & -2.841  & 36.26 & 7.549 & 51.45 & 7.722  & 52.47	& 7.653	& 48.11 & 7.783  & 38.23  & 7.658 & 46.4   & 7.77  & 39.17  &  7.689 \\    
            & 6247.557 & 3.892 & -2.435  & 50.89 & 7.459 & 78.65 & 7.848  & 76.92	& 7.696	& 67.16 & 7.755  & 58.06  & 7.673 & 68.38  & 7.815 & 59.82  &  7.727 \\    
            & 7222.394 & 3.889 & -3.402  & 18.76 & 7.63  & 24.49 & 7.668  & 31.20	& 7.762	& --	& --	 & 22.18  & 7.797 & 27.26  & 7.867 & 21.07  &  7.770 \\        
            & 7449.335 & 3.889 & -3.488  & 18.06 & 7.686 & 32.46 & 7.930  & 33.91	& 7.896	& 30.34 & 8.011  & 23.90  & 7.924 & 26.7   & 7.931 & 22.45  &  7.889 \\    
            & 7711.723 & 3.903 & -2.683  & 44.82 & 7.533 & 66.54 & 7.784  & 72.21	& 7.784	& 59.00 & 7.789  & 53.25  & 7.776 & 59.31  & 7.821 & 50.48  &  7.735 \\ 

\ion{Co}{1} & 5301.039 & 1.71  & -2.0    & 19.9  & 4.945 & --    & --     & --   & --    & --    & --    & --    & --    & 15.26 & 5.017 & 24.35 & 5.054 \\
            & 6093.143 & 1.74  & -2.44   & 10.5  & 5.034 & --    & --     & --   & --    & --    & --    & --    & --    & 7.81  & 5.108 & 12.56 & 5.115 \\ 
			& 6814.942 & 1.956 & -1.9    & 18.5  & 4.959 & 11.96 & 5.07   & 13.57 & 4.956& 26.79 & 5.242 & 25.16 & 5.103 & 17.65 & 5.136 & 26.59 & 5.156 \\
			& 5647.234 & 2.28  & -1.56   & 14.06 & 4.863 & 9.34  & 4.976  & --    & --   & --    & --    & --    & --    & 11.07 & 4.935 & 19.91 & 5.042 \\
			& 6632.433 & 2.28  & -2.0    & 8.7   & 5.012 & --    & --     & --    & --   & 12.44 & 5.262 & 12.04 & 5.146 & --    & --    & 13.42 & 5.217 \\

\ion{Ni}{1} & 6643.629 & 1.676 & -2.3    & 92.89 & 6.366 & 86.9  & 6.55   & --    & --   & --    & --    & --    & --    & 91.31 & 6.48  & 108.6 & 6.636 \\
			& 6108.107 & 1.676 & -2.45   & 65.59 & 6.062 & 57.99 & 6.226  & 58.99 & 6.056& 72.29 & 6.216 & 76.1  & 6.217 & 60.95 & 6.142 & 74.99 & 6.234 \\
			& 6327.593 & 1.676 & -3.15   & 38.63 & 6.253 & 30.41 & 6.41   & 30.53 & 6.231& 45.0  & 6.431 & 49.24 & 6.418 & 35.71 & 6.385 & 49.6  & 6.449 \\
			& 5846.986 & 1.676 & -3.21   & 23.29 & 6.017 & 19.29 & 6.236  & 20.53 & 6.09 & 28.38 & 6.21  & 30.43 & 6.157 & 22.39 & 6.188 & 33.95 & 6.251 \\  
			& 5748.346 & 1.676 & -3.26   & 28.42 & 6.193 & 22.36 & 6.373  & --    & --   & 33.67 & 6.374 & --    & --    & 24.43 & 6.294 & 36.84 & 6.364 \\   
			& 6128.963 & 1.676 & -3.33   & 25.97 & 6.184 & 20.33 & 6.367  & 20.26 & 6.186& 31.53 & 6.377 & 34.14 & 6.334 & 21.23 & 6.261 & 36.57 & 6.403 \\ 
			& 6767.768 & 1.826 & -2.17   & 78.94 & 6.122 & 74.29 & 6.325  & --    & --   & --    & --    & --    & --    & 80.83 & 6.3   & 93.74 & 6.385 \\   
			& 6177.236 & 1.826 & -3.5    & 13.77 & 6.154 & 10.98 & 6.358  & --    & --   & 17.26 & 6.347 & 19.21 & 6.31  & 11.82 & 6.269 & 20.82 & 6.373 \\  
			& 5754.655 & 1.935 & -2.33   & 74.12 & 6.395 & 68.12 & 6.564  & 73.16 & 6.475& 85.24 & 6.627 & 83.04 & 6.527 & 73.38 & 6.525 & 86.08 & 6.628 \\
			& 6370.341 & 3.542 & -1.94   & 13.37 & 6.251 & 12.66 & 6.439  & --    & --   & 20.53 & 6.531 & 16.51 & 6.353 & 12.52 & 6.347 & 17.84 & 6.407 \\   
			& 6842.035 & 3.658 & -1.48   & 23.78 & 6.196 & 22.79 & 6.377  & 23.75 & 6.279& 32.77 & 6.44  & 29.41 & 6.315 & 24.41 & 6.331 & 33.59 & 6.418 \\ 
			& 6176.807 & 4.088 & -0.26   & 62.48 & 6.161 & 64.01 & 6.372  & 64.73 & 6.277& 74.79 & 6.398 & 71.56 & 6.305 & 66.65 & 6.324 & 75.28 & 6.395 \\
			& 6111.066 & 4.088 & -0.87   & 34.04 & 6.257 & 35.09 & 6.461  & 38.58 & 6.421& 42.8  & 6.462 & 40.78 & 6.38  & 37.48 & 6.427 & 42.84 & 6.434 \\
			& 6204.6   & 4.088 & -1.1    & 21.49 & 6.198 & 22.58 & 6.41   & 23.23 & 6.319& 28.58 & 6.409 & 28.4  & 6.357 & 23.98 & 6.368 & 30.63 & 6.419 \\
			& 6175.36  & 4.089 & -0.559  & 48.4  & 6.219 & 51.43 & 6.455  & 57.24 & 6.45 & 62.76 & 6.501 & 58.01 & 6.379 & 53.47 & 6.405 & 58.49 & 6.409 \\
			& 5760.828 & 4.105 & -0.8    & 33.33 & 6.203 & 32.7  & 6.373  & 36.88 & 6.349& 45.33 & 6.468 & 40.81 & 6.34  & 35.42 & 6.347 & 44.12 & 6.418 \\
			& 6223.981 & 4.105 & -0.91   & 29.89 & 6.223 & 29.3  & 6.392  & 29.42 & 6.287& 37.3  & 6.411 & 35.86 & 6.337 & 30.4  & 6.337 & 38.29 & 6.4   \\
			& 6186.709 & 4.105 & -0.96   & 31.48 & 6.308 & 30.73 & 6.474  & --    & --   & --    & --    & --    & --    & 32.7  & 6.436 & 39.76 & 6.479 \\   
			& 6230.09  & 4.105 & -1.26   & 20.34 & 6.342 & 21.26 & 6.55   & 23.48 & 6.499& 25.19 & 6.507 & 27.29 & 6.507 & 22.03 & 6.494 & 29.4  & 6.568 \\
			& 6378.247 & 4.154 & -0.83   & 31.93 & 6.226 & 33.22 & 6.432  & 32.3  & 6.307& 40.78 & 6.435 & 37.67 & 6.333 & 33.31 & 6.355 & 42.51 & 6.44  \\
			& 5805.213 & 4.167 & -0.64   & 40.8  & 6.245 & 40.52 & 6.419  & 41.17 & 6.326& 47.93 & 6.41  & 49.31 & 6.391 & 42.14 & 6.368 & 51.83 & 6.455 \\ 
			& 6598.593 & 4.236 & -0.98   & 24.83 & 6.288 & 25.5  & 6.479  & 26.81 & 6.406& 35.44 & 6.553 & 33.6  & 6.473 & 26.67 & 6.431 & 35.27 & 6.52  \\
			& 6130.13  & 4.266 & -0.96   & 22.71 & 6.262 & 21.08 & 6.394  & 20.97 & 6.289& 30.91 & 6.489 & 27.64 & 6.373 & 23.01 & 6.37  & 30.98 & 6.459 \\
			& 6635.118 & 4.419 & -0.82   & 24.34 & 6.288 & 23.84 & 6.442  & 22.09 & 6.3  & 32.79 & 6.511 & 31.49 & 6.443 & 25.61 & 6.413 & 32.79 & 6.482 \\  

\ion{Cu}{1} & 5782.127 & 1.642 & -1.72   & 77.4  & 4.46  & 63.08 & 4.481  & 67.35 & 4.368& 83.97 & 4.605 & 88.92 & 4.64 & 68.61  & 4.442 & 88.83 & 4.686 \\

\ion{Zn}{1} & 4722.153 & 4.03  & -0.338  & 69.25 & 4.463 & 74.14 & 4.615  & --    & --   & --    & --    & --    & --   & 74.11  & 4.56  & 74.89 & 4.641 \\
            
\enddata
\tablecomments{This table is published in its entirety in the online journal.  
A portion is shown here for guidance regarding 
its form and content.}
\end{deluxetable}

\begin{deluxetable}{ccccccccccc}
\tablecolumns{11}
\tablewidth{0pt}
\tabletypesize{\scriptsize}
%\tabletypesize{\tiny}
\tablecaption{Synthesized HFS Line Abundances \label{tab:hfs_lines}}
\tablehead{
     \colhead{}&
     \colhead{$\lambda$}&
     \colhead{$\chi$}&
     \colhead{}&
     \colhead{Solar}\\
     \colhead{Ion}&
     \colhead{(\AA)}&
     \colhead{(eV)}&
     \colhead{$\log \mathrm{gf}$}&
     \colhead{$\log N_{\odot}$}&
     \colhead{Pr0201}&
     \colhead{Pr0133}&
     \colhead{Pr0208}&
     \colhead{Pr0081}&
     \colhead{Pr0051}&
     \colhead{Pr0076}
	}
\startdata
\ion{Co}{1} & 6814.942 & 1.956 & -1.9 & -- &   5.07 & --   & --   & --   & --   & -- \\
            & 5301.039 & 1.71  & -2.0 & -- & --   & --   & --   & --   & --   & -- \\
       
\ion{Mn}{1} & 5432.546 & 0.0 & -3.795 & 5.28 & 5.46 & --   & --   & --   & 5.41 & 5.55 \\      
            						
\ion{Sc}{2} & 6604.601 & 1.357 &  -1.309 & 3.10 & 3.18 & 3.09 & 3.25 & 3.15 & 3.17 & 3.27 \\        
            & 6245.637 & 1.507 &  -1.03  & 3.10 & 3.23 & 3.09 & 3.25 & 3.15 & 3.16 & 3.17 \\

\ion{V}{1}  & 6111.645 & 1.043 & -0.715 & --   & --   & --   & --   & --   & --    & 4.05 \\
            & 6090.214 & 1.081 & -0.062 & 3.83 & 4.03 & --   & --   & --   & --    & 4.05 \\ 
\enddata
\tablecomments{Listed synthesized abundances replace EW abundances from Table \ref{tab:linelist}.}
\end{deluxetable}

\subsection{User Guided Equivalent Width Measurements}
The EW measurement of absorption lines is performed by an in-house user guided code called eXtract from SPECTra - Equivalent Widths (XSpect-EW), specifically created for this purpose.
The general process for measuring EWs of absorption lines for abundance derivations includes three main steps: normalization of the continuum, wavelength shift, and fitting the absorption lines with Gaussian or Voigt profiles to determine the EW of the lines.
Each step of the process can be vulnerable to user error if done manually, as detailed in the following subsections, depending on how much care and time is given to each part of the analysis.
Ideally, one would take extreme care in each of these parts within a minimal amount of time.

\subsubsection{Continuum Normalization}
Arguably the most important part of the EW measuring process is the ability to consistently determine the continuum of the spectrum or each spectral order.
The differential Curve of Growth (CoG) abundance determination involves comparing the abundances of the star of interest to those of a standard or reference star, often the Sun; therefore, one must be able to determine the continuum in each spectral order in the same way for the star and the standard so as to minimize differences in the abundances that could arise as a result of the analysis.
XSpect-EW has been designed to normalize the spectrum efficiently and accurately by using a two-step process. 

Step 1: a spectrum is split into smaller pieces (referred to as selection windows), the size of which is set to larger than the typical size of the absorption lines to ensure that the selection windows do not fall entirely within a line. 
For example, in an optical high resolution Keck/HIRES spectrum with R$>$60k, the width of a typical line will be 0.4--0.6 \AA, so the size of the selection window could be $\sim$1.6 \AA.
Points above 90\% of the flux values within a selection window are chosen as points in the continuum (90\% value can be adjusted by the user if needed).

\begin{figure}[!ht]
\centering
\includegraphics[width=0.9\linewidth,trim=0 10 0 20,clip]{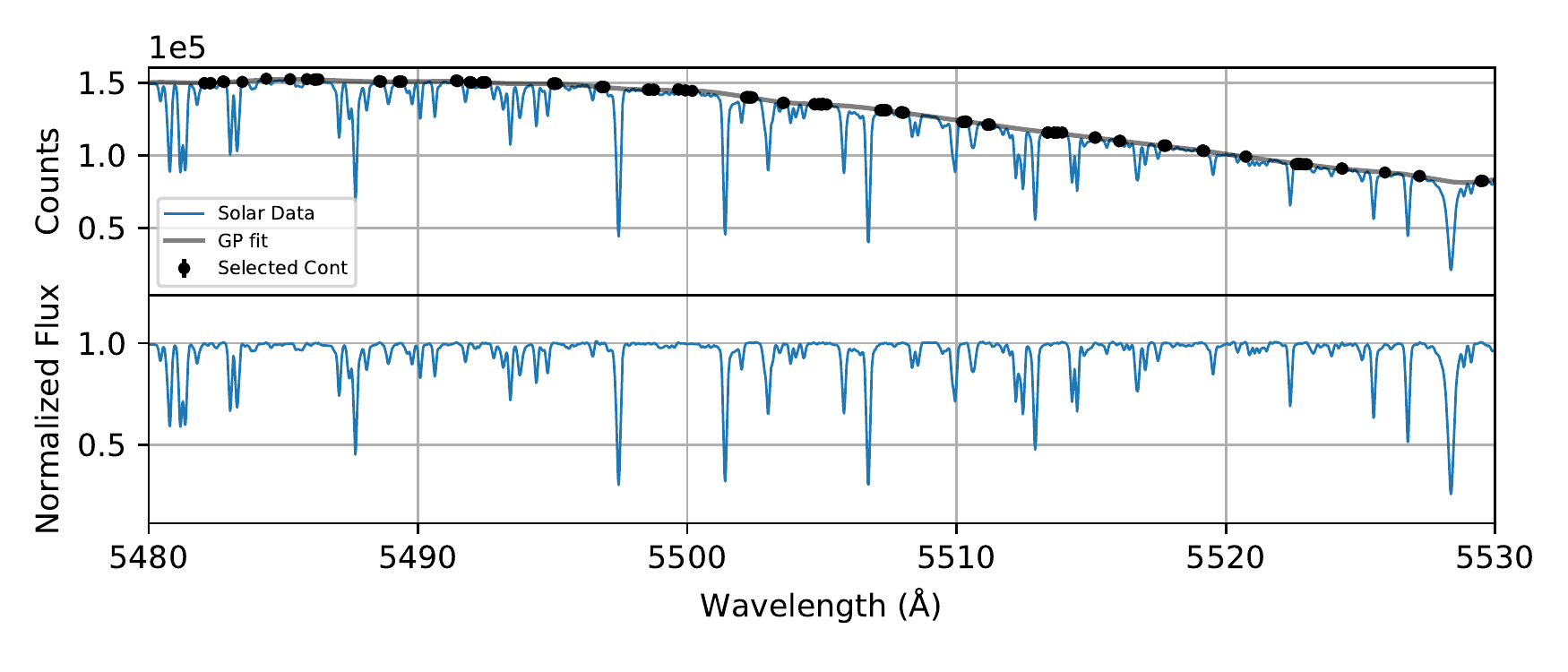}
\caption{Sample section of an order within the observed solar spectrum before and after normalization. Black points are those selected as part of the continuum; black curve is the output of the Gaussian Process.}
\label{fig:norm}
\end{figure}

Step 2: the chosen continuum points are used to normalize the spectrum by fitting them with a Gaussian Process (GP).
A GP is a collection of random variables, any finite number of which have a joint Gaussian distribution.
GPs are useful for data containing non-trivial stochastic signals or noise.
A square exponential kernel (or Gaussian kernel) $k_{SE}(x,x') = A\exp[-\Gamma(x-x')^{2}]$ is used for the covariance function in the GP where $A$ is the variance and $\Gamma$ is the inverse lengthscale ($\Gamma = (1/2l^{2})$).
A variance of similar order of magnitude as the flux values is used ($10^{7}$), along with an inverse lengthscale of 0.1.
After testing various values, these gave the best results for our spectra.
XSpect-EW will be updated to automatically determine the variance and lengthscale values for each order in a spectrum and to have a variety of kernels available.
The flux is then divided by the curve output by the GP (grey curve in Fig. \ref{fig:norm} top panel) which normalizes the order (Fig. \ref{fig:norm} bottom panel).

\subsubsection{Wavelength Shift}
Once normalization is complete, each order is wavelength shifted to the rest frame. 
This can be done manually by visually identifying a portion of an order in the observed spectrum, comparing it to a reference spectrum, and using software to shift the wavelength axis by the appropriate amount. 
Errors in this part of the process can lead to measuring the wrong lines within an order, which may show up as abundance outliers or result in higher than expected abundance errors later in the analysis.

XSpect-EW makes use of a simple $\chi^{2}$ minimization approach to this problem. 
An input spectrum is shifted to match a solar spectrum order by order. 
This is done by determining the median wavelength value in the order of the spectrum to be shifted, finding which order in the solar spectrum this value falls, and shifting the order to match the solar spectrum.
The shift is determined by evaluating $\chi^{2}$ between both orders for a range of wavelength shifts from $-$5 \AA\ to +5 \AA\ in steps of 0.1 \AA\ and selecting the shift with the minimum $\chi^{2}$ value (shift ranges and resolution can be changed by user if needed). 
In cases where a shift cannot be determined for an order, the shifts for other orders are used to predict the shift of the missing order by linear interpolation of the shift values. 
A user may also manually shift any order by a specified amount for further correction if needed.

\subsubsection{Equivalent Width Measurements}
Measuring the EW of an absorption line generally involves fitting a Gaussian or Voigt function to the line and then integrating that function to determine the area enclosed by the curve and continuum. XSpect-EW uses a four step process for this.

\begin{figure}[h!]
    \center
    \includegraphics[width=4.2in,trim=0 0 0 30,clip]{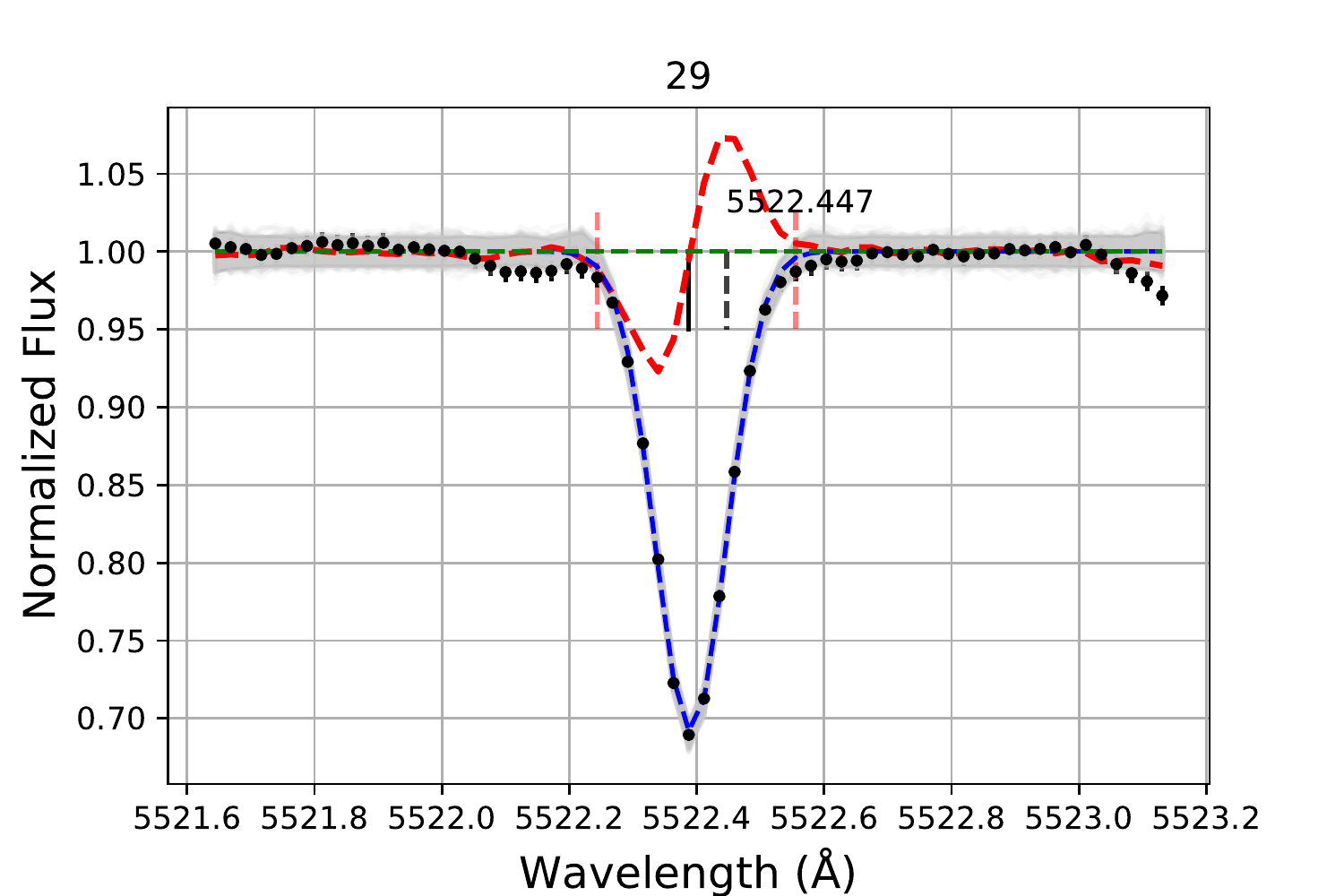}
    \caption{Sample plot for an Fe I line (5522.447\AA) from the solar spectrum. Black dots with error bars correspond to data; grey shading shows the $1\sigma$ variance, blue curve is the Gaussian fit. Green horizontal line defines the continuum. Red curve shows the slope of the flux values (shifted to the continuum). Red vertical lines establish the boundaries of the line measured. Dashed black vertical line is the input wavelength of the line, solid vertical line is the best fit wavelength of the line.}
    \label{fig:EW}
\end{figure}

Step 1: the extent of the line being measured is determined. Carefully determining the extent, or width of the line at the continuum, is critical to obtaining an accurate measure of the absorption in the line.
A piece of the spectrum around a selected line is trimmed from the rest of the spectrum (default set to 1.5 \AA\ about the line center). 
The boundaries of the line being measured are determined by utilizing the slope of the flux values within the piece of the spectrum. 
These values range from a maximum or minimum near the center of the line to zero at the continuum and the core, as shown in Figure~\ref{fig:EW}. 
Using one half of the standard deviation of the flux slope values, a range of values centered on the continuum is created.
After testing various values, one half of the standard deviation gave us best results relative to hand-measured EWs.
The boundaries of the absorption line are defined as the points where the slope enters this range from the maximum or minimum values to either edge of the line.
At this boundary, the slope values approach zero as the flux values approach the constant continuum. 

Step 2: with the boundaries defined, XSpect-EW assumes the continuum is correct and shifts the data outside the boundaries of the line (still within the 1.5 \AA\ line window) to match the continuum, effectively isolating the line to be measured.

Step 3: the isolated absorption line and associated continuum is fed into another GP (same kernel) with a small variance of 1 and large inverse lengthscale of 100.

Step 4: the GP can produce different realizations of the data, and XSpect-EW uses this to perform a Gaussian fit to the line, repeating the process 100 times and thus producing 100 different EW measurements. 
The mean and standard deviation are calculated using the 100 measurements of the line. 
Each absorption line can be plotted and remeasured if needed, with the ability to adjust the local continuum, center of the line, and boundaries of the line.

\begin{figure}
    \centering
    \includegraphics[width=5.5in]{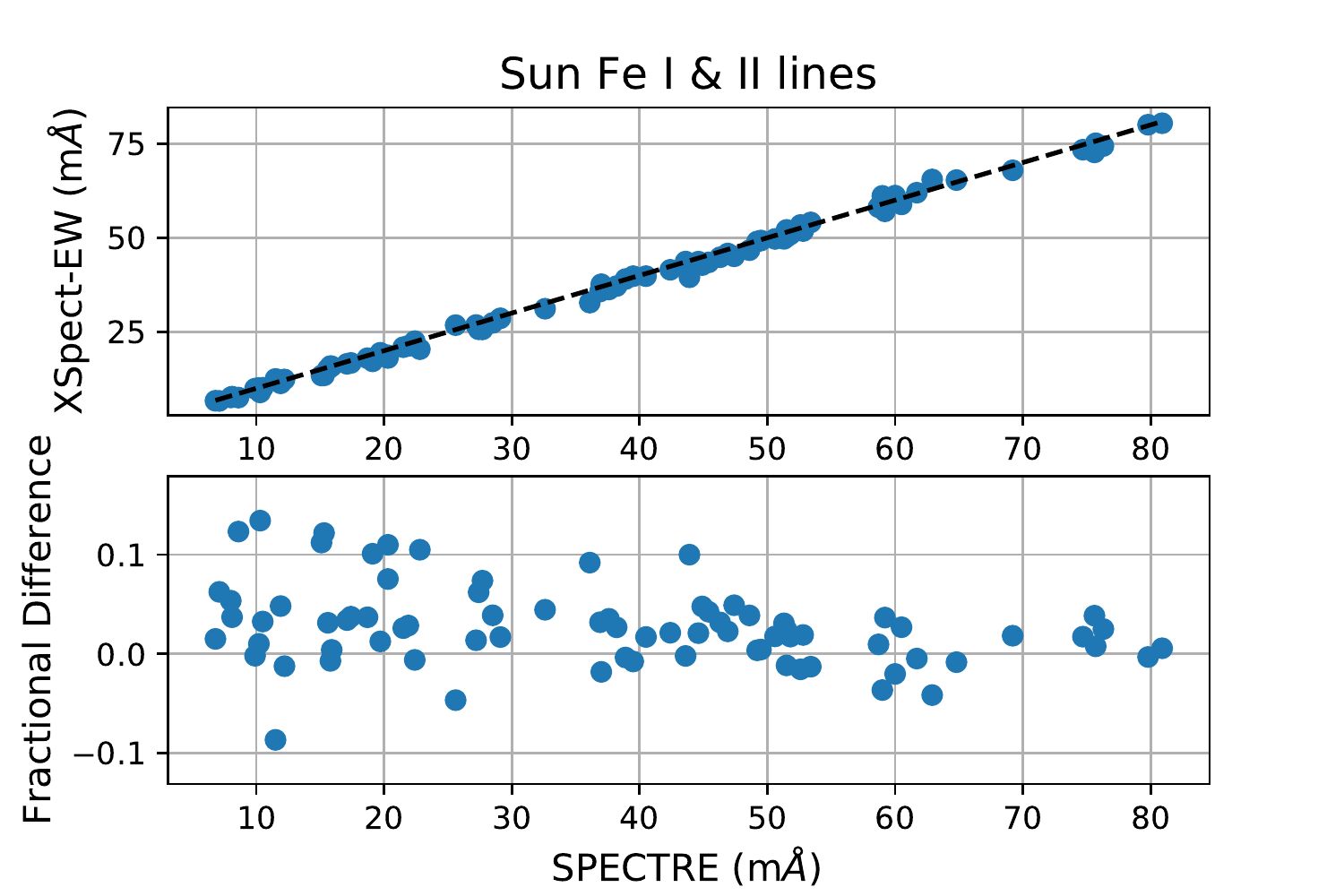}
    \caption{Top panel: XSpect-EW measured lines (with some user input) plotted against hand measured EWs in SPECTRE. Dashed line is the one-to-one line. Bottom panel: Fractional difference ($\left[\mathrm{XSpect{\text -}EW} - \mathrm{hand}\right] /$ hand) between XSpect-EW and hand measured values.}
    \label{fig:EW_comparison}
\end{figure}

\subsubsection{Verification of Methodology}
\begin{figure}
    \centering
    \includegraphics[width=5.5in]{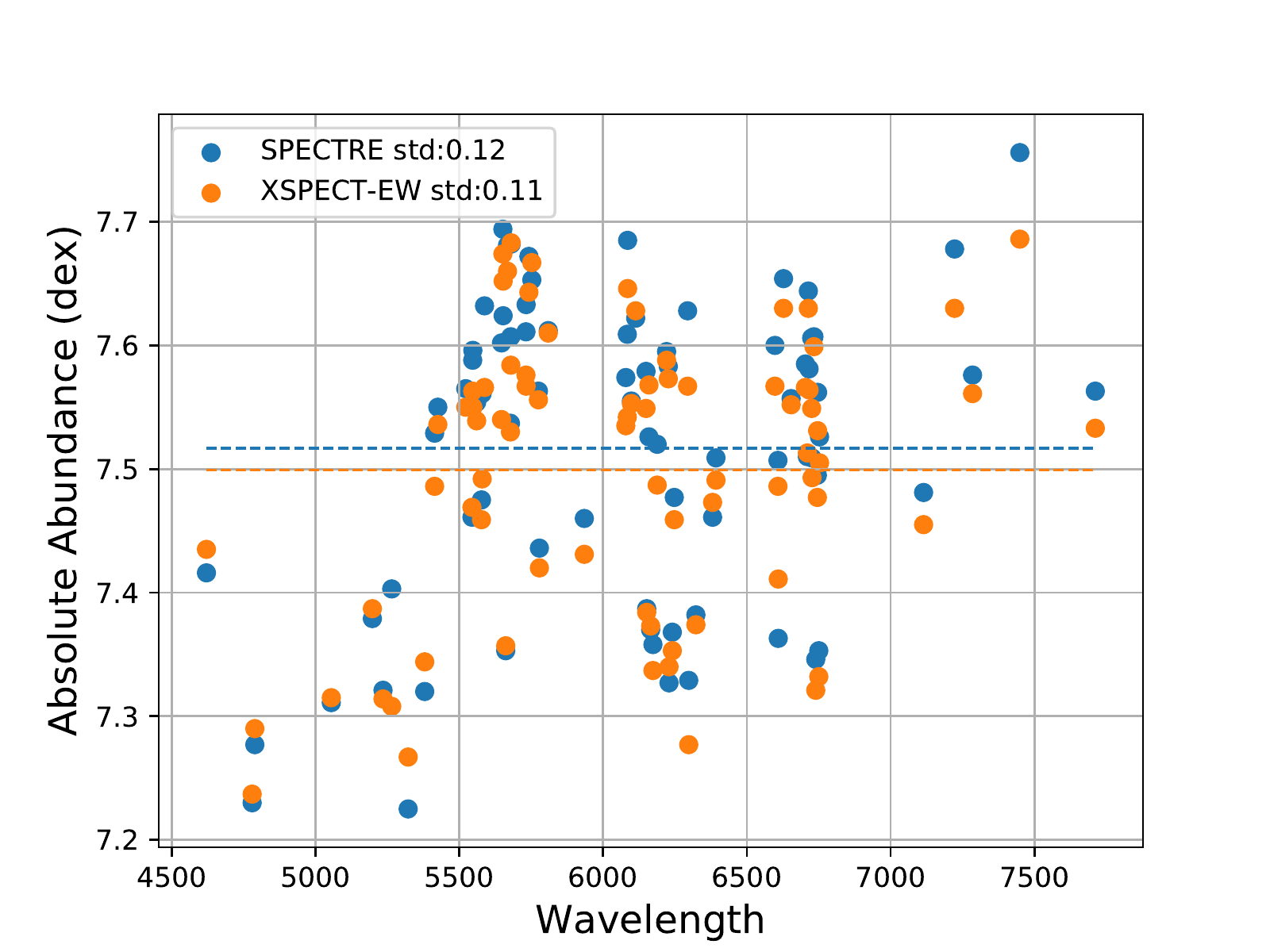}
    \caption{Output absolute abundances from Solar \ion{Fe}{1} and \ion{Fe}{2} lines shown in Figure \ref{fig:EW_comparison} using MOOG. Blue points represent abundances from SPECTRE measurements while orange points represent abundances from XSpect-EW. The dashed lines (same colors used as for points) show the average abundances for each method.}
    \label{fig:abs_abund_comparison}
\end{figure}

To verify the robustness of our new EW measuring tool, in Figure \ref{fig:EW_comparison} we compare EWs of \ion{Fe}{1} and \ion{Fe}{2} lines measured in our solar spectrum with XSpect-EW and by hand with SPECTRE, a spectral analysis package \citep{2012ascl.soft02010S}. All differences are $<$ 15\%, with 80\% of Fe lines having differences smaller than 5\%. In the remaining elemental lines (not shown here), $\sim$70\% of lines have differences smaller than 5\%. XSpect-EW requires some user input to obtain these measurements which consists of checking each automatic line fit and remeasuring problematic lines by adjusting extra parameters that characterize the absorption line. We note that this interactive functionality has been purposely built into XSpect-EW for this specific purpose. The scatter in the fractional difference of the line measurements is larger for the other elements, because those lines can generally be more difficult to measure due to the variety in the strength of the lines and proximity to other lines.
A well curated line-list can be helpful for automatic line fits as strong, isolated absorption lines are easier for XSpect-EW to measure automatically.
The quality of the data will also be a deciding factor in the number of lines that need to be remeasured. In this example, 17\% of Fe lines and 23\% of other elemental lines required remeasuring, greatly reducing the amount of time needed to measure all lines and allowing the user to focus on problematic lines to produce abundances with high precision.
Lines that are remeasured are only considered based on visual inspection of the fit, not in comparison to previous measurements, since when measuring a new star no comparison would be available. The average of the Fe absolute abundances derived from the XSpect-EW and SPECTRE EWs, shown in Figure \ref{fig:abs_abund_comparison}, agree within errors and agree with the input solar absolute abundance within MOOG of 7.50 dex. For all Fe I and II lines, we see a smaller scatter in the absolute abundances from the XSpect-EW measurements than in the SPECTRE measurements by 0.01 dex. From this we can see that XSpect-EW performs as well, if not slightly better, than a full set of hand-measured EWs from SPECTRE.

From the \ion{Fe}{1} and \ion{Fe}{2} abundances derived using the EWs measured with XSpect-EW, we determine an average cluster metallicity for Praesepe of $0.21\pm 0.02$ dex, which is in good agreement with most past works on the cluster metallicity.
In Table \ref{tab:compare_parameters}, we show the literature stellar parameter values for each star along with our own derived values. 
Our derived stellar parameters ($\rm T_{\rm eff}, \log G, [Fe/H]$) are also in good agreement with current literature values.
The main discrepancy between our stellar parameters and literature values comes from the $\xi$ parameter values which in general are larger than literature values, likely due to the higher $\xi$ value adopted for the Sun in this work (1.38 km/s here, compared to lower values used in other studies)\citep{2008AA...489..403P, 2019ApJ...871..142G, 2020AA...633A..38D}.
A more detailed comparison of individual stars to the literature is presented in section \ref{s:mean_compare}.

\begin{deluxetable}{lcccccc}
\tablecolumns{6}
\tablewidth{0pt}
%\rotate
%\tabletypesize{\scriptsize}
%\tabletypesize{\tiny}
\tablecaption{Comparison to Literature\label{tab:compare_parameters}}
\tablehead{
\colhead{Star ID }& \colhead{$\rm T_{\rm eff}$}& \colhead{$\rm \log G$}& \colhead{$\rm \xi$}& \colhead{$\rm [Fe/H]$}& \colhead{Reference}& \\
\colhead{alt ID} & \colhead{$\rm K$} & \colhead{$\rm \log(g\cdot cm\cdot sec^{-1})$} & \colhead{$\rm km/s$} & \colhead{$\rm dex$} &  \\}

\startdata

Pr0201  & $6168\pm35$  & $4.34\pm0.10$ & $1.52\pm0.06$ & $0.23\pm0.05$  & This work \\
Prae kw 418  &  $6174\pm50$ & $4.41\pm0.10$ & ---  & $0.19\pm0.04$ & \citet{2012ApJ...756L..33Q} \\
            & $6062\pm110$ & $4.44\pm0.07$ & $1.27\pm0.18$ & $0.24\pm0.10$ &  \citet{2008AA...489..403P} \\
\hline
Pr0133  &$6067\pm60$&$4.18\pm0.12$&$1.74\pm0.11$&$0.19\pm0.06$& This work \\ 
Prae kw 208  & $6005\pm19$& $4.46\pm0.21$& $1.05\pm0.04$& $0.18\pm0.03$& \citet{2019ApJ...871..142G}\\
            & $5997\pm60$ & $4.38\pm0.20$ & $1.40\pm0.20$ & $0.12\pm0.10$ & \citet{2013ApJ...775...58B} \\
            & $5993\pm110$ & $4.45\pm0.07$ & $1.52\pm0.18$ & $0.28\pm0.10$ & \citet{2008AA...489..403P} \\
\hline

Pr0208  &$5869\pm46$&$4.37\pm0.13$&$1.53\pm0.07$&$0.26\pm0.07$& This work \\
N2632-8 & $5977\pm75$& $4.55\pm0.15$&$1.30\pm0.20$&$0.25\pm0.11$& \citet{2020AA...633A..38D} \\
Prae kw 432 & $5841\pm73$ & $4.40\pm0.20$ & $1.25\pm0.20$ & $0.17\pm0.10$ & \citet{2013ApJ...775...58B} \\
\hline

Pr0081  &$5731\pm42$&$4.44\pm0.11$&$1.41\pm0.06$&$0.18\pm0.06$& This work \\
CPrae kw 30 & $5716\pm45$ & $4.57\pm0.42$ & $1.18\pm0.04$ & $0.12\pm0.04$ & \citet{2019ApJ...871..142G} \\
            & $5675\pm111$ & $4.44\pm0.20$ & $1.07\pm0.20$ & $0.12\pm0.10$ & \citet{2013ApJ...775...58B} \\
\hline

Pr0051  &$6017\pm27$&$4.40\pm0.07$&$1.54\pm0.05$&$0.16\pm0.03$& This work \\
TYC 1395-668-1 &&&&& \\
\hline

Pr0076  &$5748\pm24$&$4.44\pm0.07$&$1.35\pm0.03$&$0.22\pm0.05$& This work \\
Prae kw 23& $5773\pm53$ & $4.56\pm0.24$ & $1.20\pm0.04$ & $0.20\pm0.04$ & \citet{2019ApJ...871..142G} \\
            &$5699\pm79$ & $4.43\pm0.20$ & $1.10\pm0.20$ & $0.12\pm0.10$ & \citet{2013ApJ...775...58B} \\
            
\enddata
\end{deluxetable}

\section{Results\label{s:results}}

\subsection{Derived Stellar Abundances}
The derived elemental abundances for our targets are summarized in Table \ref{tab:params}.
%The line lists, EWs, and $\log(N)$ line-by-line abundances for each element are given in Table \ref{tab:linelist}.
Elements with only one measured absorption line have no mean and therefore no uncertainty in the mean; for those lines we have adopted the total error to be the average total error of all other elements in the star.
Some elements in some stars were not measurable due to lack of spectral coverage or removal of lines that were not measurable (due to noise, blending, bad fit...etc).
The errors for all stellar parameters and abundances are symmetric or close to symmetric (where $\pm\sigma_{Total}$ intervals are equal or close to equal); in all cases we conservatively adopt the larger error, except with surface gravity where we adopt the average error.
The resulting differences between the archive data (2003) and our data (2013) for Pr0076 are within error bars for the derived stellar parameters and elemental abundances, as shown in Table \ref{tab:params}.
Despite the fact that we are using data of differing quality, observing conditions, instrument setup, and time of observation we are able to produce consistent results giving us great confidence in our analysis.

\begin{figure}[H]
\centering
\includegraphics[width=\linewidth,trim=0 10 0 10,clip]{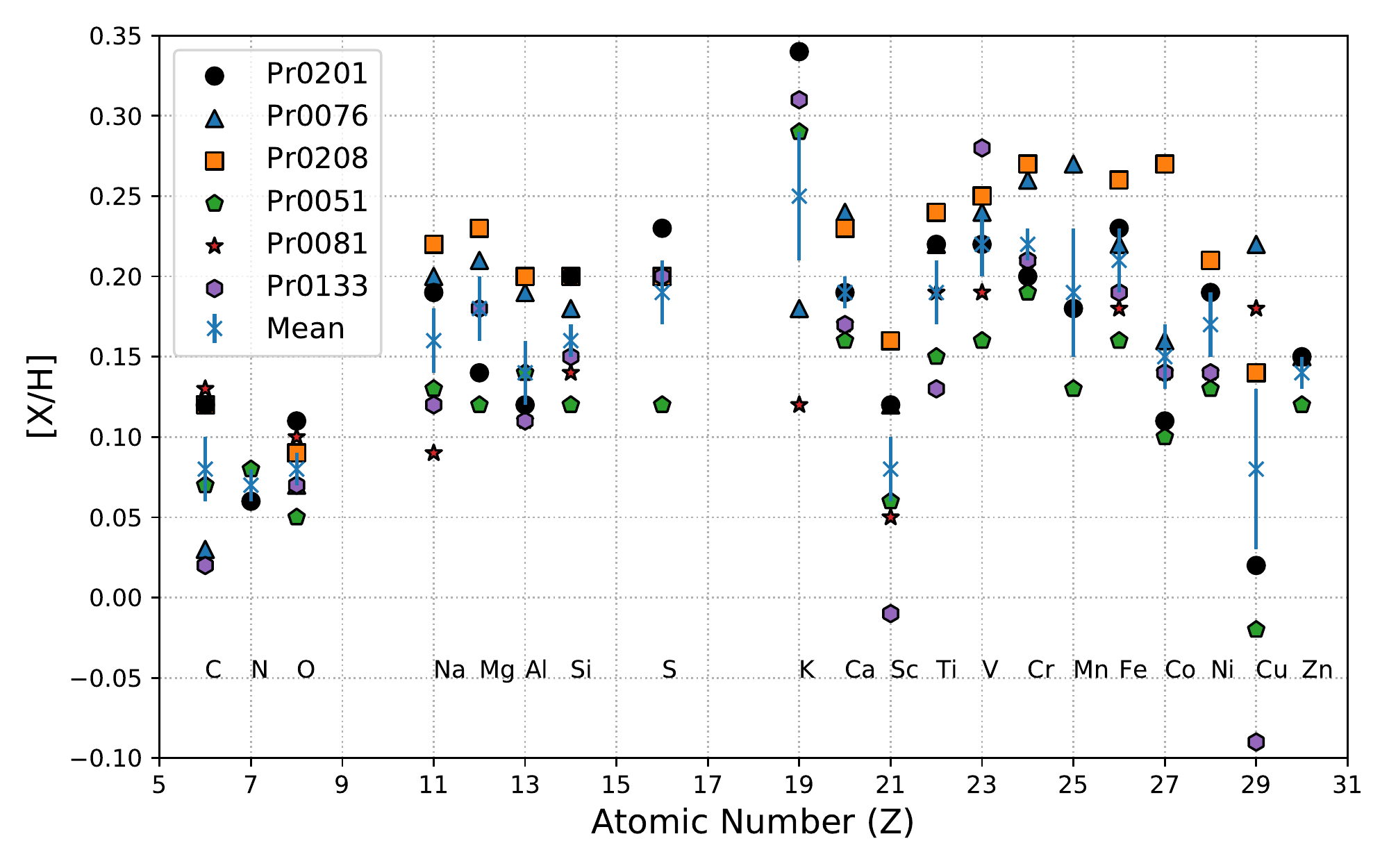}
\caption{Abundances vs.\ atomic number (Z) for all Praesepe stars (2013 data used for Pr0076). The planet host, Pr0201, is shown as a filled circle. The typical error bar for elements is 0.05 dex (check Table \ref{tab:params} for specific errors). The mean for each element is shown in the blue cross along with the error in the mean.}
\label{fig:abd_vs_z}
\end{figure}

Fig \ref{fig:abd_vs_z} shows the abundance vs atomic number for each star and the cluster mean. 
Most elements have a similar spread about the mean while some (K, Sc, Co, and Cu) show a larger spread.
This larger spread is likely due to the fact that these elements have a low number of lines measured, lines may be weak, noisy, or blended with other nearby lines making them difficult to measure, or some combination of these factors. 
%For example in Sc we might synthesize two lines (6604.601\AA\ and 6245.637\AA) and end up with two or three total measured lines for a star where the synthesized lines may or may not always agree with non-synthesized lines.
%Other notable elements with either one or two measured lines are N, Al, S, Mn, and Zn.
%Below we verify our results with the literature, take a closer look at the individual differences between stars and the cluster, and mention notable trends.

%====================
%ABUNDS TABLE  (REVISED)
%====================
\begin{deluxetable}{lccccccc}
\tablecolumns{7}
\tablewidth{0pt}
\rotate
\tabletypesize{\scriptsize}
%\tabletypesize{\tiny}
\tablecaption{Stellar Abundances\label{tab:params}}
\tablehead{
	\colhead{}&
	\colhead{Pr0201}&
	\colhead{Pr0133}&
	\colhead{Pr0208}&
	\colhead{Pr0081}&
	\colhead{Pr0051}&
	\colhead{Pr0076 (2003)}&
	\colhead{Pr0076 (2013)}
	}
\startdata
%\teff (K)           & $6168\pm35$   & $6067\pm60$   & $5869\pm46$   & $5731\pm42$ 	& $6017\pm27$   & $5789\pm72$     & $5748\pm24$\\
%$\log g$ (cgs)      & $4.34\pm0.10$  & $4.18\pm0.12$ & $4.37\pm0.13$ & $4.44\pm0.11$ & $4.40\pm0.07$ & $4.48\pm0.12$   & $4.44\pm0.07$\\
%$\xi$ (km s$^{-1}$) & $1.52\pm0.06$ & $1.74\pm0.11$ & $1.53\pm0.07$ & $1.41\pm0.06$ & $1.54\pm0.05$ & $1.33\pm0.04$   & $1.35\pm0.03$\\
$[$C/H$]$  & $0.12\pm0.03\tablenotemark{b}\,\pm0.05\tablenotemark{c}$ & $0.02\pm0.03\,\pm0.06$ & $0.12\pm0.06\,\pm0.09$ & $0.13\pm0.03\,\pm0.05$ & $0.07\pm0.03\,\pm0.04$ & $0.10\pm0.07\,\pm0.09$ & $0.03\pm0.01\,\pm0.04$\\
$[$O/H$]$  & $0.11\pm0.03\,\pm0.05$ & $0.07\pm0.03\,\pm0.07$ & $0.09\pm0.05\,\pm0.08$ & $0.10\pm0.09\,\pm0.11$ & $0.05\pm0.03\,\pm0.05$ & ...        & $0.07\pm0.04\,\pm0.05$\\
$[$Na/H$]$ & $0.19\pm0.02\,\pm0.03$ & $0.12\pm0.04\,\pm0.05$ & $0.22\pm0.04\,\pm0.03$ & $0.09\pm0.03\,\pm0.04$ & $0.13\pm0.01\,\pm0.02$ & $0.27\pm \rm ...\,\pm0.08$ & $0.20\pm0.03\,\pm0.03$\\
$[$Mg/H$]$ & $0.14\pm0.01\,\pm0.03$ & $0.18\pm \rm ...\,\pm0.05$ & $0.23\pm0.03\,\pm0.05$ & ...   & $0.12\pm0.01\,\pm0.02$ & $0.20\pm0.24\,\pm0.24$ & $0.21\pm0.01\,\pm0.05$\\
$[$Al/H$]$ & $0.12\pm \rm ...\,\pm 0.05$& $0.11\pm \rm ...\,\pm0.05$ & $0.20\pm0.00\,\pm0.02$ & $0.11\pm0.01\,\pm0.02$ & $0.14\pm0.00\,\pm0.01$ & ...        & $0.19\pm0.02\,\pm0.03$\\
$[$Si/H$]$ & $0.20\pm0.01\,\pm0.01$ & $0.15\pm0.01\,\pm0.02$ & $0.20\pm0.01\,\pm0.01$ & $0.14\pm0.01\,\pm0.01$ & $0.12\pm0.01\,\pm0.01$ & $0.19\pm0.02\,\pm0.02$ & $0.18\pm0.01\,\pm0.01$\\
$[$Ca/H$]$ & $0.19\pm0.02\,\pm0.03$ & $0.17\pm0.01\,\pm0.05$ & $0.23\pm0.02\,\pm0.05$ & $0.17\pm0.02\,\pm0.04$ & $0.16\pm0.01\,\pm0.03$ & $0.27\pm \rm ...\,\pm0.08$ & $0.24\pm0.01\,\pm0.03$\\
$[$Sc/H$]$ & $0.12\pm0.02\,\pm0.05$ & $-0.01\pm \rm...\,\pm0.05$ & $0.16\pm0.01\,\pm0.07$ & $0.05\pm \rm...\,\pm0.05$ & $0.06\pm0.01\,\pm0.03$ & ... & $0.12\pm0.05\,\pm0.07$\\
$[$Ti/H$]$ & $0.22\pm \rm 0.01\,\pm 0.06$& $0.13\pm0.01\,\pm0.06$ & $0.24\pm0.01\,\pm0.04$ & $0.19\pm0.03\,\pm0.05$ & $0.15\pm0.01\,\pm0.04$ & $0.24\pm0.03\,\pm0.10$ & $0.22\pm0.02\,\pm0.06$\\
$[$V/H$]$  & $0.22\pm0.03\,\pm0.05$ & $0.28\pm0.05\,\pm0.08$ & $0.25\pm0.02\,\pm0.06$ & $0.19\pm0.02\,\pm0.05$ & $0.16\pm0.01\,\pm0.03$ & $0.27\pm0.03\,\pm0.08$ & $0.24\pm0.01\,\pm0.03$\\
$[$Cr/H$]$ & $0.20\pm0.01\,\pm0.03$ & $0.21\pm0.01\,\pm0.05$ & $0.27\pm0.03\,\pm0.05$ & $0.21\pm0.02\,\pm0.04$ & $0.19\pm0.01\,\pm0.02$ & $0.29\pm0.01\,\pm0.05$ & $0.26\pm0.01\,\pm0.02$\\
$[$Mn/H$]$ & $0.20\pm \rm ... \,\pm0.04$ & ...                    & ...                    & ...                    & $0.13\pm \rm ... \,\pm0.03$ & $0.36\pm0.01\,\pm0.07$ & $0.27\pm \rm ...\,\pm0.04$\\
$[$Fe/H$]$ & $0.23\pm0.01\,\pm0.05$ & $0.19\pm0.01\,\pm0.06$ & $0.26\pm0.01\,\pm0.07$ & $0.18\pm0.01\,\pm0.06$ & $0.16\pm0.00\,\pm0.03$ & $0.25\pm0.01\,\pm0.06$ & $0.22\pm0.00\,\pm0.05$\\
$[$Co/H$]$ & $0.11\pm0.01\,\pm0.03$ & $0.14\pm \rm ...\,\pm0.05$ & $0.27\pm0.02\,\pm0.04$ & $0.14\pm 0.01\,\pm0.03$	& $0.10\pm0.03\,\pm0.04$ & $0.18\pm0.03\,\pm0.07$ & $0.16\pm0.02\,\pm0.03$\\
$[$Ni/H$]$ & $0.19\pm0.01\,\pm0.02$ & $0.14\pm0.01\,\pm0.04$ & $0.21\pm0.01\,\pm0.03$ & $0.14\pm0.01\,\pm0.03$ & $0.13\pm0.01\,\pm0.02$ & $0.24\pm0.01\,\pm0.05$ & $0.21\pm0.01\,\pm0.02$\\
$[$Zn/H$]$ & $0.15\pm \rm ...\,\pm 0.05$& ...                & ...                    & ...                    & $0.12\pm \rm ... \,\pm0.03$ & $0.18\pm \rm ... \,\pm0.08$ & $0.15\pm \rm ...\,\pm0.04$\\
$[$S/H$]$  & $0.23\pm0.09\,\pm0.09$ & $0.20\pm \rm ...\,\pm0.05$  &$0.20\pm \rm ...\,\pm0.05$& ...             & $0.12\pm0.02\,\pm0.03$ & $0.22\pm0.02\,\pm0.05$ & $0.20\pm0.03\,\pm0.05$\\
$[$Cu/H$]$ & $0.02\pm \rm ...\,\pm 0.05$& $-0.09\pm \rm ...\,\pm0.05$ &$0.14\pm \rm ...\,\pm0.05$& $0.18\pm \rm ...\,\pm0.05$    & $-0.02\pm \rm ...\,\pm0.03$ & $0.28\pm \rm ...\,\pm0.08$ & $0.22\pm \rm ...\,\pm0.04$\\
$[$K/H$]$  & $0.34\pm \rm ...\,\pm 0.05$& $0.31\pm \rm ...\,\pm0.05$  & ...                       & $0.12\pm \rm ...\,\pm0.05$	& $0.29\pm \rm ...\,\pm0.03$ & ... & $0.18\pm \rm ...\,\pm0.04$\\
$[$N/H$]$  & $0.06\pm \rm ...\,\pm 0.05$& ...                          & ...                       & ...	& $0.08\pm \rm ...\,\pm0.03$ & ... & ...\\
\enddata
%\tablenotetext{a}{Adopted solar parameters: \teff\ $=5777$ K, $\log g=4.44$, and $\xi=1.38$ km s$^{-1}$.}
\tablenotetext{b}{$\sigma_{\mu}$-- the uncertainty in the mean}
\tablenotetext{c}{$\sigma_{Total}$-- quadratic sum of $\sigma_{\mu}$ and uncertainties due to \teff, $\log g$, and $\xi$.}
%\tablenotetext{d}{In both stars, the abundance measurements for Mg and Mn were determined from a single spectral line.  That is why the uncertainty in the mean is 0.00 in both stars.}
\end{deluxetable}

\subsubsection{Planet-Hosting Star}
\begin{figure}[!ht]
\centering
\includegraphics[width=\linewidth,trim=10 0 10 30,clip]{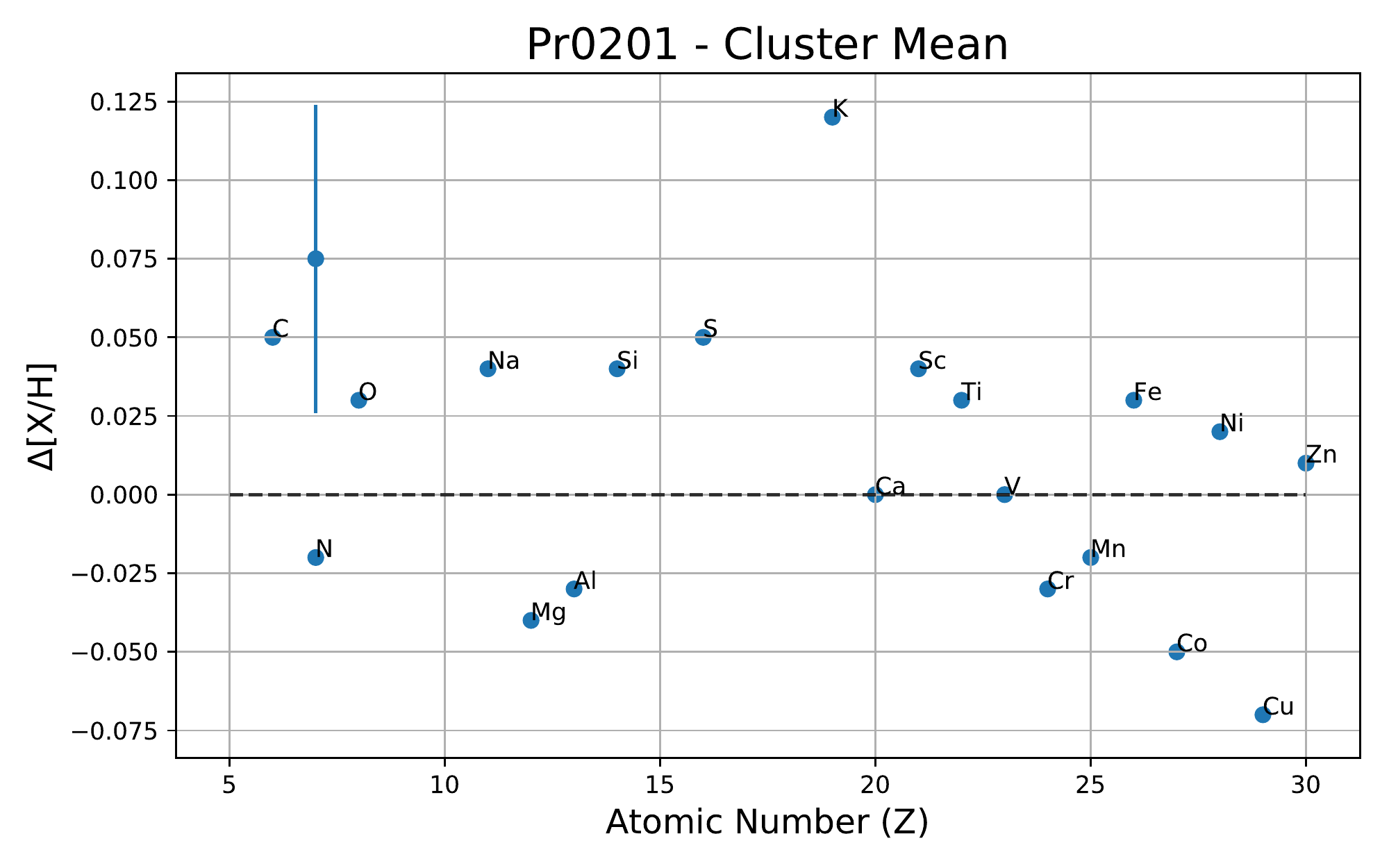}
\caption{Abundance difference between planet host (Pr0201) and cluster mean vs.\ atomic number for all elements. Dashed line denotes zero difference and the average error is shown in the top left corner, which shows the $\pm1\sigma$ error with a dot to mark the center.}
\label{fig:host_mean_num_all}
\end{figure}

In order to see if the planet host, Pr0201, is different from the rest of the our sample, we compare the abundances to the cluster mean. 
Figure \ref{fig:host_mean_num_all} shows the difference between Pr0201 and the cluster mean (not including the planet host) for each derived element vs atomic number.
Most elements fall between $\pm 0.05$ dex of the cluster mean and are within errors of zero with a few exceptions.
The element with the largest difference is K having an overabundance of 0.12 dex when compared to the cluster.
Si, having the smallest total error of 0.01 dex, is also significantly overabundant in the planet host by 0.04 dex.
The planet host shows no obvious trend in abundance vs atomic number over all measured elements.

\subsubsection{Non Planet-Hosting Stars}
\begin{figure}[!ht]
\centering
\includegraphics[width = 3.2 in]{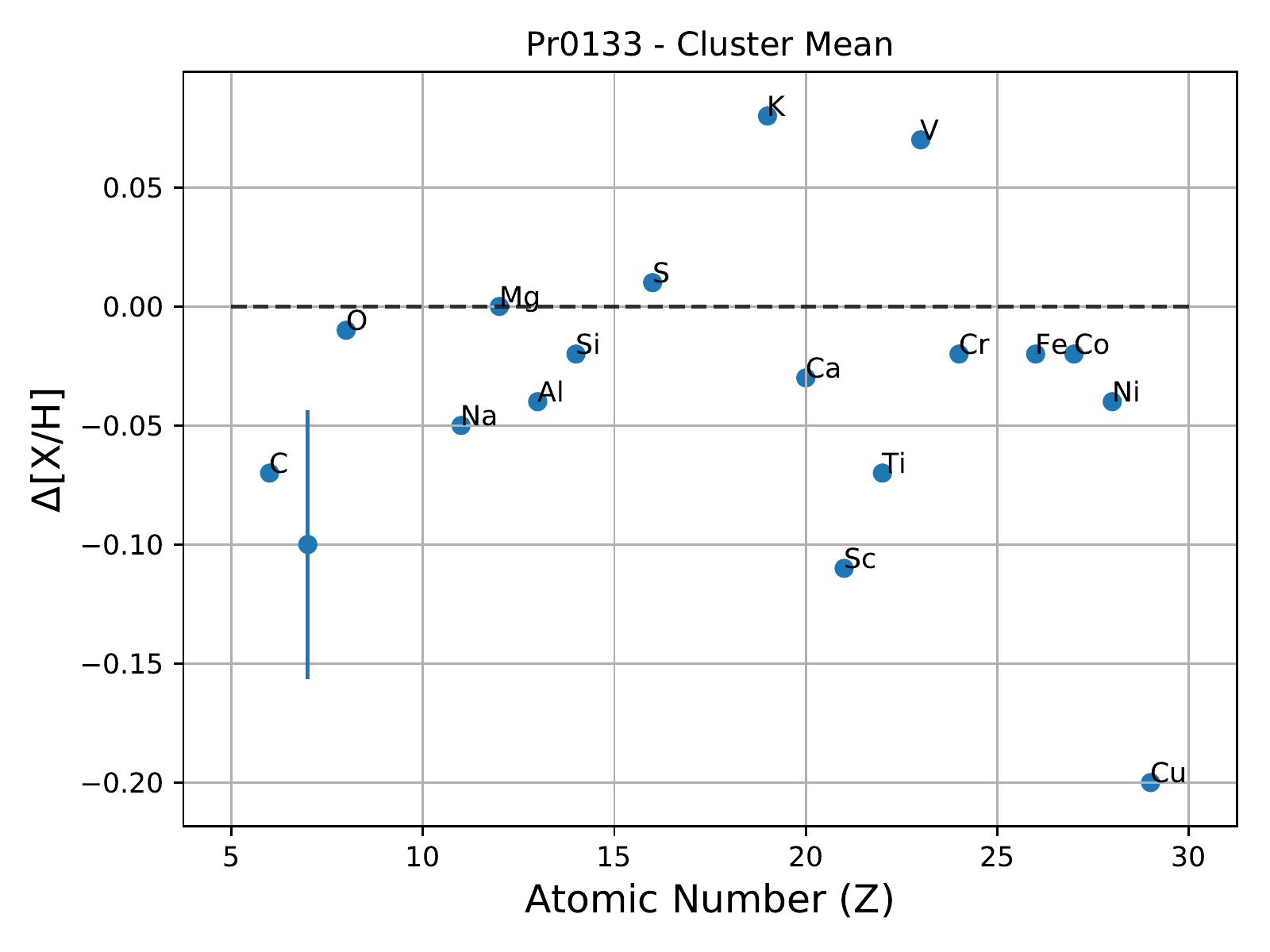}
\includegraphics[width = 3.2 in]{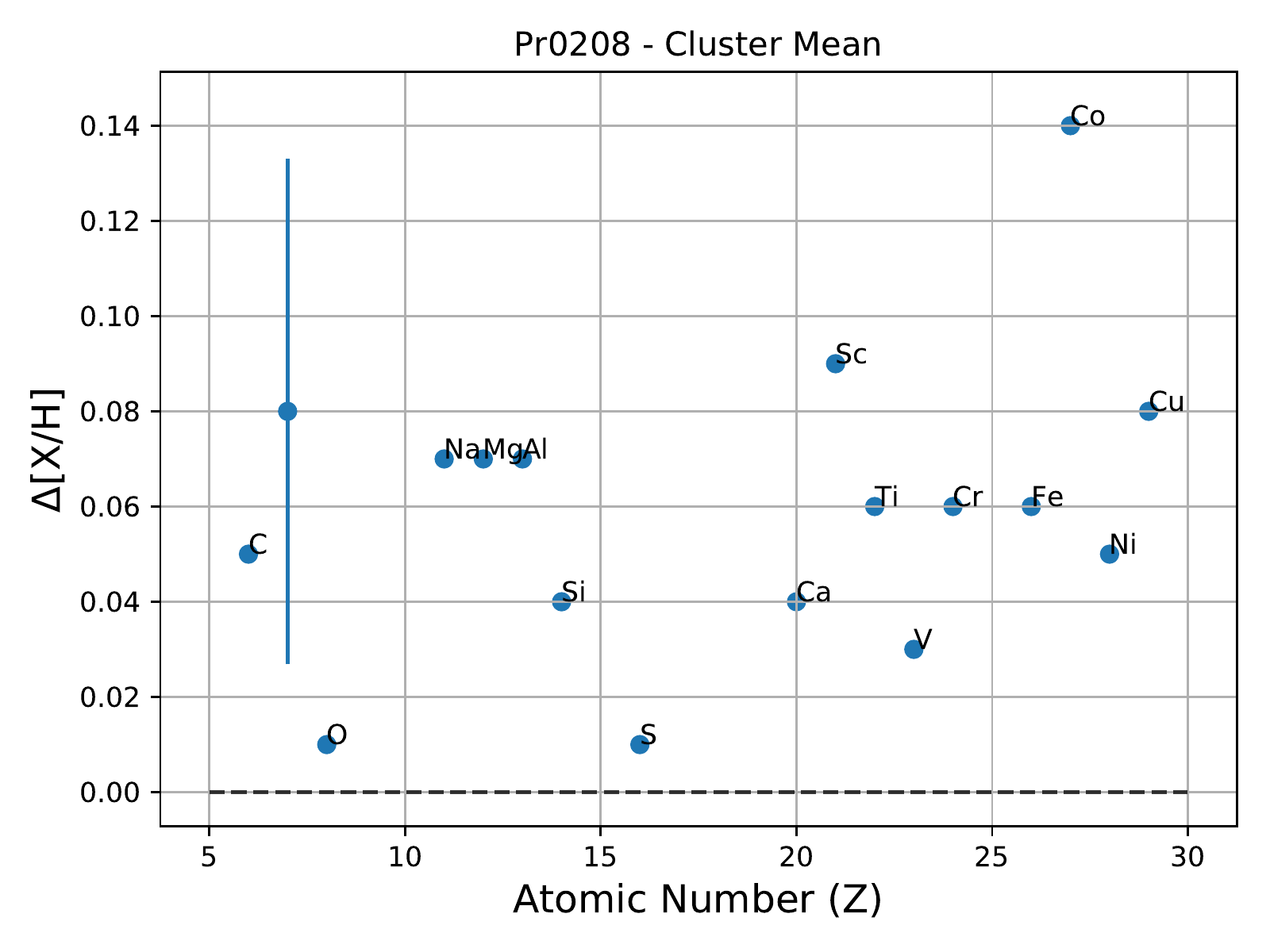}
\includegraphics[width = 3.2 in]{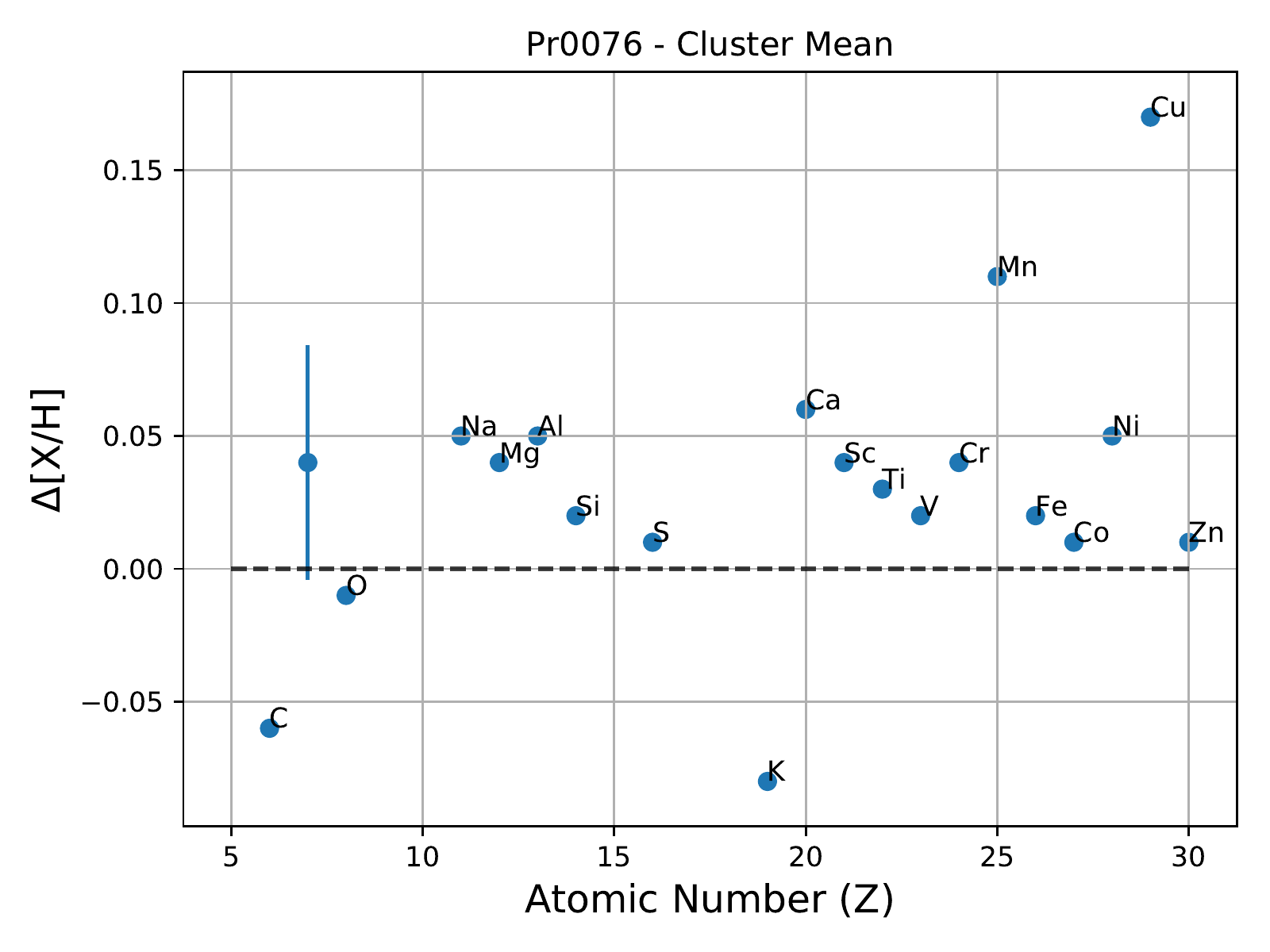}
\includegraphics[width = 3.2 in]{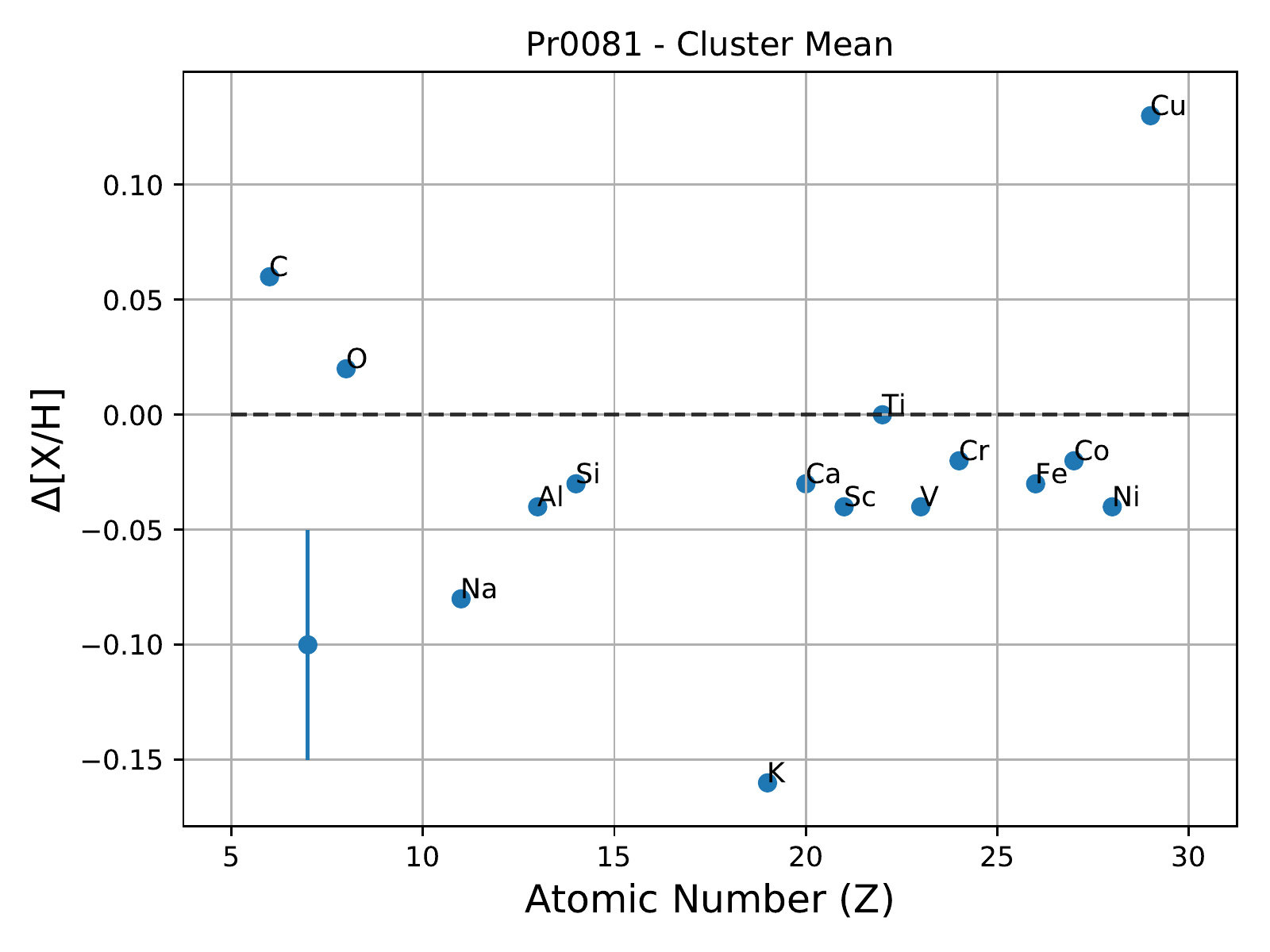}
\includegraphics[width = 3.2 in]{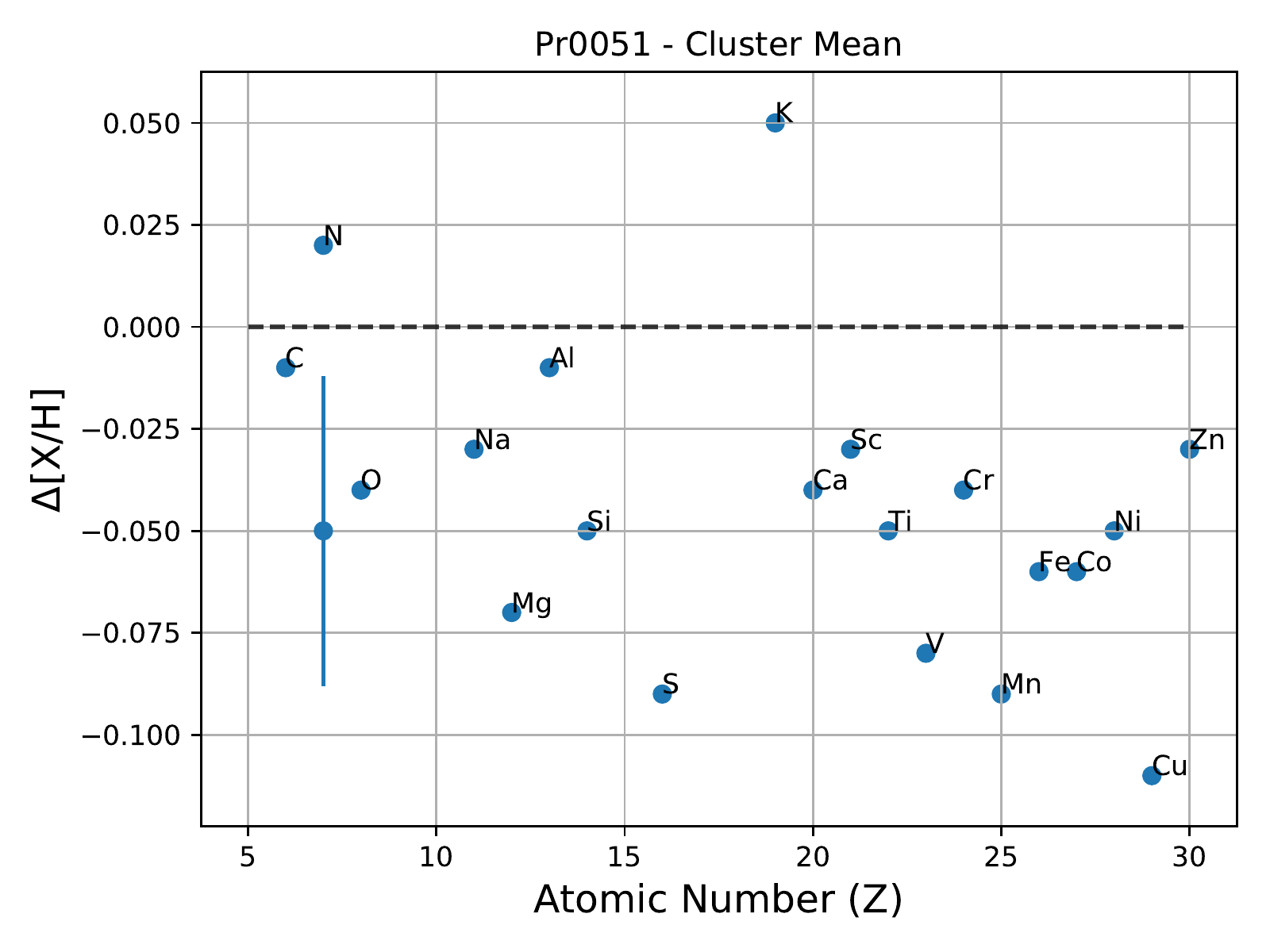}
\caption{Abundance difference between each other star in Praesepe and cluster mean vs.\ atomic number for all elements. Dashed line denotes zero difference and the average error is shown on the left of each plot, which shows the $\pm1\sigma$ error with a dot to mark the center.}
\label{fig:others_mean_tc_all}
\end{figure}

In Figure \ref{fig:others_mean_tc_all} we show the difference between each star and the cluster mean (the mean not including that specific star).
Elements noted below are considered overabundant or deficient if they are more than 1$\rm \sigma$ (in actual errors, not average errors) different from zero.

Pr0133 is only overabundant in K by 0.08 dex and deficient in C by 0.07 dex, Ti by 0.07 dex, Sc by 0.11 dex, and Cu by 0.20 dex.

Pr0208 has the highest metallicity of our analysed stars, [Fe/H] of 0.26 dex, and shows an overabundance in many elements: S by 0.04 dex, Ni by 0.05 dex, Ti by 0.06 dex, Cr by 0.06 dex, Na by 0.07 dex, Mg by 0.07 dex, Al by 0.07 dex,  Sc by 0.09 dex, and Co by 0.14 dex.
While elements of high condensation temperature tend to be overabundant in Pr0208, the high error bars on the volatile elements make it difficult to say if there is any overall trend with condensation temperature.

Pr0081 shows an overabundance in C by 0.06 dex and Cu by 0.13 dex.
Deficient elements include: Si by 0.03 dex, Al by 0.04 dex, Na by 0.08 dex, and K by 0.16 dex.

Pr0051 shows no overabundance but many deficiencies: Cr by 0.04 dex, Si by 0.05 dex, Ni by 0.05 dex, Ti by 0.05 dex, V by 0.08 dex, Fe by 0.06 dex, Mg by 0.07 dex, Co by 0.06 dex, S by 0.09 dex, Mn by 0.09 dex, and Cu by 0.11 dex.

Pr0076 shows deficiencies in C by 0.06 dex and K by 0.08 dex while being overabundant in: Cr by 0.04 dex, Na by 0.05 dex, Al by 0.05 dex, Ni by 0.05 dex, Ca by 0.06 dex, Mn by 0.11 dex, and Cu by 0.17 dex.

\subsection{Comparison to Previous Work\label{s:mean_compare}}
\citet{2007ApJ...655..233A} determined [Fe/H] of four G type dwarfs in Praesepe through spectroscopy (equivalent width method) and photometry (photometric metallicity).
The mean [Fe/H] they determined through spectroscopy ($+0.11\pm0.03$) was lower than our determined mean metallicity of $0.21\pm0.02$.
The [Fe/H] they determined through photometry ($+0.20\pm0.04$) is in much better agreement with our results and most current literature values.

\citet{2008AA...489..403P} performed a detailed chemical study of eight elements (Fe, Na, Al, Si, Ca, Ti, Cr, Ni) on 20 solar-type stars in four open clusters, obtaining high resolution (R=$\rm100K$) and high signal to noise (S/N = 130) data for seven stars in Praesepe (with two overlapping stars; Pr0133 and Pr0201).
They measure a higher mean Fe abundance of $+0.27\pm0.10$, though still within errors of our mean abundance.
For Pr0201, all eight elements agree within errors, while Pr0133 shows less agreement.
Half of the elements agree with our results (Ca, Ti, Cr, Fe) while the other half (Na, Al, Si, Ni) are higher in abundance than our measurements.
Our Al measurement comes from just one absorption line for Pr0133 with the total error being the average of the other elements, this could mean this error is underestimated.
It is unclear why there is such a discrepancy in the other elements.
All reported mean abundances relative to Fe are within error bars except for O which is much lower than our determined abundance.

\citet{2011A&A...535A..30C} studied abundances of three giants in Praesepe using the equivalent width method and while they measure a lower mean [Fe/H] abundance of $+0.16\pm0.05$ dex, this mean does fall within errors of our determined mean.
Many of the other abundances relative to Fe (Al, Ca, Co, Cr, Mg, Na, Ni, Sc, Si, Ti, and V) disagree with our results having higher abundances except in the case of Ti and Ca.
This is likely due to the fact that these stars are of different stellar type than ours and the lower measured metallicity would increase the abundance ratios relative to Fe.

\citet{2013ApJ...775...58B} presented chemical abundances of 16 elements (Li, C, O, Na, Mg, Al, Si, Ca, Sc, Ti, V, Cr, Fe, Ni, Y, and Ba) for 11 solar-type stars in Praesepe (with four overlapping stars; Pr0133, Pr0208, Pr0081, and Pr0076) through equivalent width analysis.
They determined a mean cluster metallicity of $+0.12 \pm 0.04$ dex, which is lower than our determined mean by $0.09$ dex.
Abundance ratios for other elements relative to Fe are within $1\sigma$ of our measurements except for Al and Sc.
In most cases the abundance ratios are higher in value, likely because of the lower mean [Fe/H] abundance.
In Pr0133, Pr0208, and Pr0076, our study overlaps with 13 elements (excluding Li, Y, and Ba) while Pr0081 overlaps with 12 elements (excluding Mg as well).
For Pr0133, all elements agree within errors except for Sc, which was found to be about solar or higher (0.04$\pm$0.08 dex) while we derived a sub-solar value (-0.20$\pm$0.08 dex).
For Pr0208, errors are not reported in this study, but most elements agree within our own erros.
Two elements that do not agree within our errors are Sc and Ti; however, if the errors on these elements are at all similar to the errors for the other stars reported in \citet{2013ApJ...775...58B} then it is very likely they would agree with our abundances.
Abundances for Pr0081 are also presented without errors and similarly to Pr0208, all derived abundances (Fe, C, O, Na, Al, Si, Ca, Sc, Ti, V, Cr, and Ni) are within our own errors except for Sc.
It is also true that if the errors for this star are similar to the other stars, Sc would also fall within the errors.
For Pr0076, all 13 elements fall within reported errors.

\citet{2019ApJ...871..142G} used an automated spectral analysis code (BACCHUS) to determine abundances of 24 elements for five stars in Praesepe (with three overlapping stars; Pr0133, Pr0081, and Pr0076).
They report mean abundances for G and K stars separately, $+0.17\pm0.04$ and $+0.12\pm0.01$ dex respectively. These means are consistent with each other and the literature, the average for G types is consistent with our work.
In Pr0133 and Pr0081, our study overlaps with 14 elements (C, O, Na, Mg, Al, Si, Ca, Sc, Ti, V, Cr, Co, Ni, and Cu), all of which agree within errors for Pr0081.
For Pr0133, all elements but Na and Cu agree within errors.
Our [Na/Fe] abundance is about solar or less (-0.07$\pm$0.07 dex) while \citet{2019ApJ...871..142G} determined an even lower abundance of -0.35$\pm$0.16 dex.
The reverse is true for Cu, our work derived a very low abundance of -0.28$\pm$0.09 dex while \citet{2019ApJ...871..142G} derived a solar or higher value of 0.02$\pm$0.03 dex.
For Pr0076, our study overlaps with 17 elements (Fe, C, O, Na, Mg, Al, Si, Ca, Sc, Ti, V, Cr, Mn, Co, Ni, Cu, Zn), all of which also fall within reported errors.

\citet{2020AA...633A..38D} revisited the metallicity of Praesepe taking high-resolution spectroscopic observations of 10 solar-type dwarf stars, including Pr0208.
All eight elements (Fe, Na, Mg, Al, Si, Ca, Ti, and Ni) studied are within errors of our derived abundances for Pr0208.
They report a mean metallicity of $+0.21\pm0.01$ dex and conclude that Praesepe is the most metal-rich, young open cluster in the solar neighborhood, in remarkable agreement with our results that also gives confidence in works that report a higher metallicity.
Pr0051 has no literature values to be compared to but the general agreement of our results to the literature gives us confidence in those values as well.

Even with similar analyses the literature suggests a large scatter in metallicity for Praesepe, however, each individual work shows a much lower scatter within their respective sample.
The scatter within these works (for [X/Fe]), including our own, is largely consistent with what we would expect from open clusters \citep{2019A&A...628A..54K,2007AJ....133.1161D,2016ApJ...817...49B}.
This implies that the Praesepe cluster is chemically homogeneous and these works may be self-consistent but they may not all be consistent with each other.
As our analyses and precision improve, it seems we may converge on a consistent metallicity for Praesepe, perhaps a larger sample analyzed consistently could shed more light on the mean and scatter for the cluster.

\subsection{Chemical Homogeneity of Praesepe}
In Table \ref{tab:abundance_scatter} we show the [X/H] and [X/Fe] abundance mean, standard deviation, and error in the mean for our stars in Praesepe.
For most of the elements, the [X/Fe] standard deviation and the error in the mean are lower than either for [X/H].
A few exceptions are the elements C, N, O, and Zn which show a lower [X/H] standard deviation and error in the mean.
K shows the same standard deviation and error in the mean in both [X/H] and [X/Fe].
The average standard deviation is $\sim$0.01 dex lower in [X/Fe] than [X/H], while the average error in the mean is about the same.
In [X/Fe], the standard deviation of most elements is at or below the average $\sim$0.02-0.03 dex, well within literature limits on open cluster abundance scatter \citep{2007AJ....133.1161D,2016ApJ...817...49B,2019A&A...628A..54K}.
With this, we verify that the stars in our sample are chemically homogeneous; a larger sample would be required to confirm the chemical homogeneity of Praesepe as a whole.
C and N are above the average by $\sim0.02$ dex.
K and Cu are much higher than the average by $\sim0.04$ and $\sim0.06$ dex respectively.
Looking at the error in the mean we see a similar separation where most elements are below the average at $\sim$0.01-0.02 dex with elements N, K, and Cu being $\sim0.03$ dex higher than the average.
C and Mn are above the mean by $\sim0.01$ dex or less.

For elements with $Z > 19$ (Ca, Sc, Ti, V, Cr, Mn, Co, Ni, Zn), the scatter in both standard deviation and error in the mean is higher for odd-Z elements (Sc, V, Mn) with the exception of Co which is similar to Ca, Ti, and Cr, but higher than Ni and Zn. 
This is likely due to the fact that some of these odd-Z elements may have 1-2 synthetically measured absorption lines for each star to account for hfs effects along with lines measured normally where hfs effects are negligible.

\begin{deluxetable}{lcccc}
\tablecolumns{4}
\tablewidth{0pt}
\tabletypesize{\footnotesize}
\tablecaption{Abundance Scatter\label{tab:abundance_scatter}}
%\caption{Mean, standard deviation, and error in the mean for all elements relative to H and Fe. Average standard deviation and error in the mean is also shown in the last row relative to H and Fe.}
\tablehead{
	\colhead{[X/Y]}&
	\colhead{Mean}&
	\colhead{Std}&
	\colhead{Error in Mean}
	}
\startdata
	$\rm [C/H]$   & $ 0.082$ & $0.045$ & $ 0.02$ \\
	$\rm [C/Fe]$  & $-0.125$ & $0.048$ & $0.021$ \\
	\hline
	$\rm [N/H]$   & $  0.07$ & $ 0.01$ & $ 0.01$ \\
	$\rm [N/Fe]$  & $-0.125$ & $0.045$ & $0.045$ \\
	\hline
	$\rm [O/H]$   & $ 0.082$ & $ 0.02$ & $0.009$ \\
	$\rm [O/Fe]$  & $-0.125$ & $0.029$ & $0.013$ \\
	\hline
	$\rm [Na/H]$  & $ 0.158$ & $0.047$ & $0.021$ \\
	$\rm [Na/Fe]$ & $-0.048$ & $0.024$ & $0.011$ \\
	\hline
	$\rm [Mg/H]$  & $ 0.176$ & $0.041$ & $0.021$ \\
	$\rm [Mg/Fe]$ & $-0.036$ & $0.029$ & $0.015$ \\
	\hline
	$\rm [Al/H]$  & $ 0.145$ & $0.037$ & $0.016$ \\
	$\rm [Al/Fe]$ & $-0.062$ & $ 0.03$ & $0.014$ \\
	\hline
	$\rm [Si/H]$  & $ 0.165$ & $ 0.03$ & $0.014$ \\
	$\rm [Si/Fe]$ & $-0.042$ & $0.009$ & $0.004$ \\
	\hline
	$\rm [S/H]$   & $  0.19$ & $0.037$ & $0.018$ \\
	$\rm [S/Fe]$  & $-0.022$ & $0.026$ & $0.013$ \\
	\hline
	$\rm [K/H]$   & $ 0.248$ & $0.084$ & $0.042$ \\
	$\rm [K/Fe]$  & $ 0.052$ & $0.084$ & $0.042$ \\
	\hline
	$\rm [Ca/H]$  & $ 0.193$ & $0.031$ & $0.014$ \\
	$\rm [Ca/Fe]$ & $-0.013$ & $ 0.02$ & $0.009$ \\
	\hline
	$\rm [Sc/H]$  & $ 0.083$ & $0.056$ & $0.025$ \\
	$\rm [Sc/Fe]$ & $-0.123$ & $0.036$ & $0.016$ \\
	\hline
	$\rm [Ti/H]$  & $ 0.189$ & $ 0.04$ & $0.018$ \\
	$\rm [Ti/Fe]$ & $-0.017$ & $0.021$ & $ 0.01$ \\
	\hline
	$\rm [V/H]$   & $ 0.223$ & $0.039$ & $0.018$ \\
	$\rm [V/Fe]$  & $0.017$ & $0.034$ & $0.015$ \\
	\hline
	$\rm [Cr/H]$  & $ 0.223$ & $ 0.03$ & $0.014$ \\
	$\rm [Cr/Fe]$ & $ 0.017$ & $0.023$ & $ 0.01$ \\
	\hline
	$\rm [Mn/H]$  & $ 0.193$ & $0.058$ & $0.041$ \\
	$\rm [Mn/Fe]$ & $ -0.01$ & $0.043$ & $0.031$ \\
	\hline
	$\rm [Fe/H]$  & $ 0.207$ & $0.033$ & $0.015$ \\
	\hline
	$\rm [Co/H]$  & $  0.153$ & $0.056$ & $ 0.025$ \\
	$\rm [Co/Fe]$ & $-0.053$ & $0.38$ & $ 0.017$ \\
	\hline
	$\rm [Ni/H]$  & $  0.17$ & $0.034$ & $0.015$ \\
	$\rm [Ni/Fe]$ & $-0.037$ & $0.014$ & $0.006$ \\
	\hline
	$\rm [Cu/H]$  & $ 0.075$ & $0.112$ & $ 0.05$ \\
	$\rm [Cu/Fe]$ & $-0.132$ & $0.104$ & $0.047$ \\
	\hline
	$\rm [Zn/H]$  & $  0.14$ & $0.014$ & $ 0.01$ \\
	$\rm [Zn/Fe]$ & $-0.063$ & $0.017$ & $0.012$ \\
	\hline
	$\rm [X/H]_{\rm avg}$ &        & $0.043$ &  $0.021$ \\
	$\rm [X/Fe]_{\rm avg}$ &       & $0.035$ &  $0.018$ \\
\enddata
\end{deluxetable}

\section{Discussion\label{s:disc}}
Differential abundances studies often compare the abundances of a planet host with that of a wide stellar companion assumed to have formed of the same cloud of material \citep{2011ApJ...740...76R,2014ApJ...790L..25T,2014ApJ...787...98M}.
The resulting differences are attributed to processes occurring after the stars have formed such as planet formation, system evolution, and stellar evolution.
Generally, open clusters are assumed to be chemically homogeneous \citep{2006AJ....131..455D,2013ApJ...775...58B} (which we verify for our sample in section \ref{s:results}), here we have the rare opportunity to compare our planet host Pr0201 with not just one stellar companion, but five other stars that formed from the same cloud with no known planets.

\subsection{Abundance Trends and Condensation Temperature}
In the search for planet formation signatures we specifically focus on trends in refractory elements with \tc\ $>$ 900K \citep{2009ApJ...704L..66M}.
During early disk evolution, elements with high \tc\ are expected to condense into solids at shorter distances from the host star, leading to refractory-poor gas and refractory-rich planetesimals. This results in two possible abundance signatures for the host star.
The removal of these elements from the protoplanetary disk allows the accretion of the refractory-depleted material onto the host star which imparts a decreasing trend in the refractory elements with \tc. Alternatively, accretion of refractory-rich planetesimals or planets themselves would impart an increasing trend.

In this section we interpret condensation temperature trends in the context of a planet engulfment model, explained in \citet{2014ApJ...787...98M}, based on the addition or removal of material with similar composition as the Earth to/from the convection zone of a solar-type star.
The composition of the Earth is taken from \citet{McDonough01} and solar composition is from \citet{2009ARA&A..47..481A}.
We adjust the size of the convection zone based on the temperature of the star according to \citet{2001ApJ...556L..59P}.
Solar abundances are modified to match those of the cluster mean (excluding Pr0201)
then we can adjust the number of \ME\ accreted/sequestered in order to produce a desired \tc\ slope.

\begin{figure}[!ht]
\centering
\includegraphics[width=\linewidth,trim=0 0 0 22,clip]{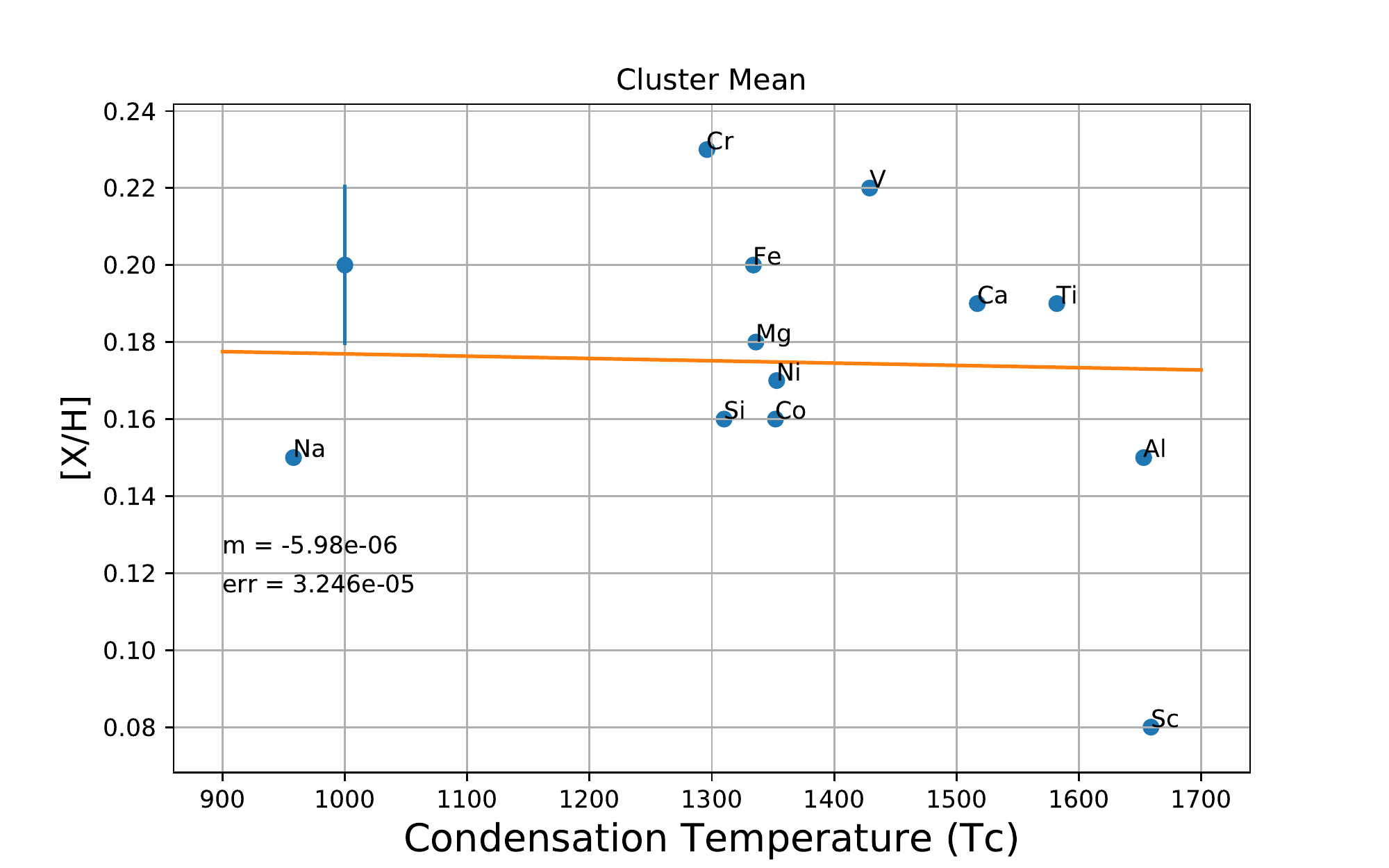}
\caption{Cluster mean vs.\ condensation temperature (\tc) for refractory elements excluding the planet host Pr0201. The average error is shown on the top left, which shows the $\pm1\sigma$ error with a dot to mark the center.}
\label{fig:mean_no_host_tc}
\end{figure}

\begin{figure}[!ht]
\centering
\includegraphics[width = 3 in]{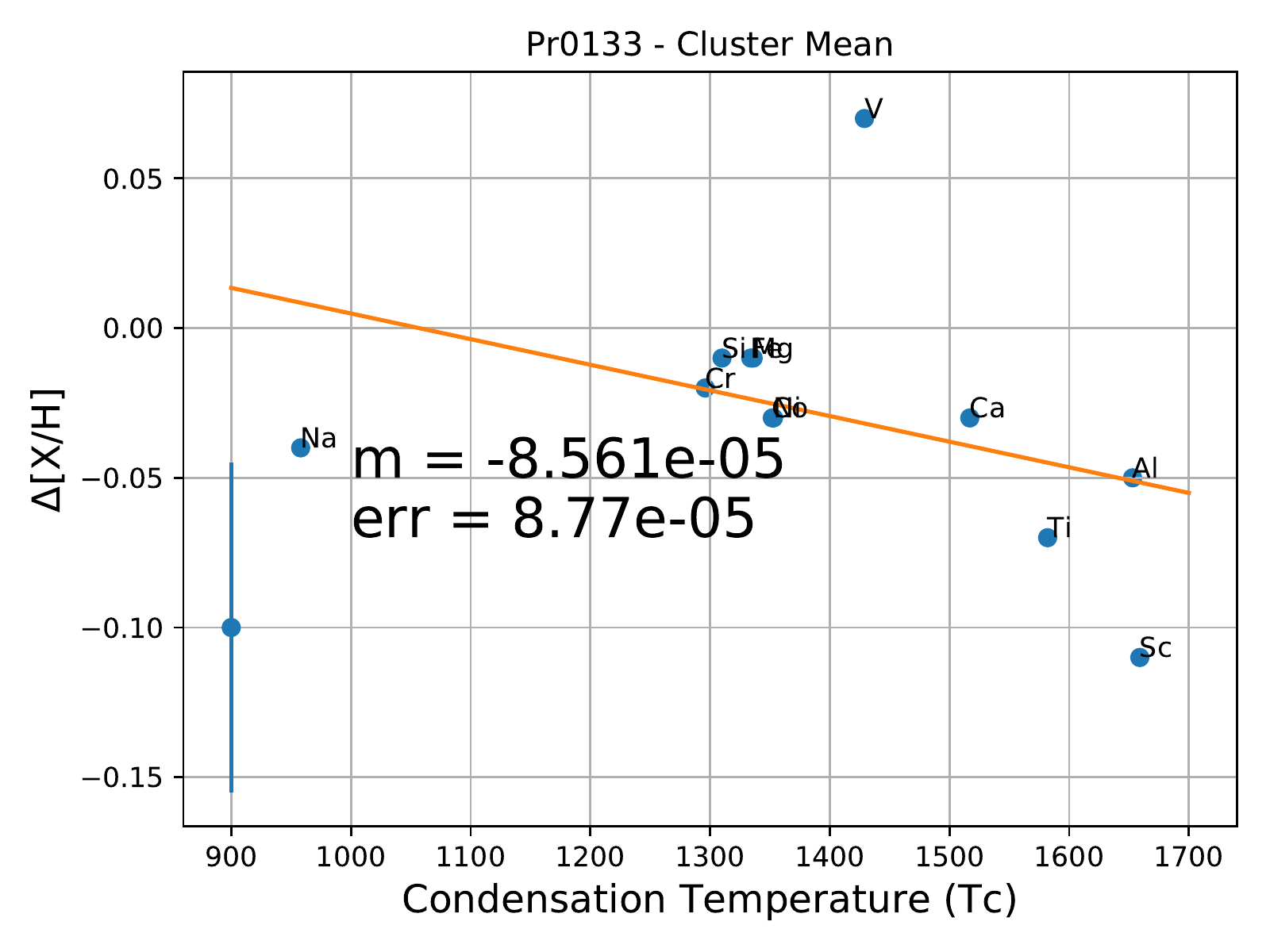}
\includegraphics[width = 3 in]{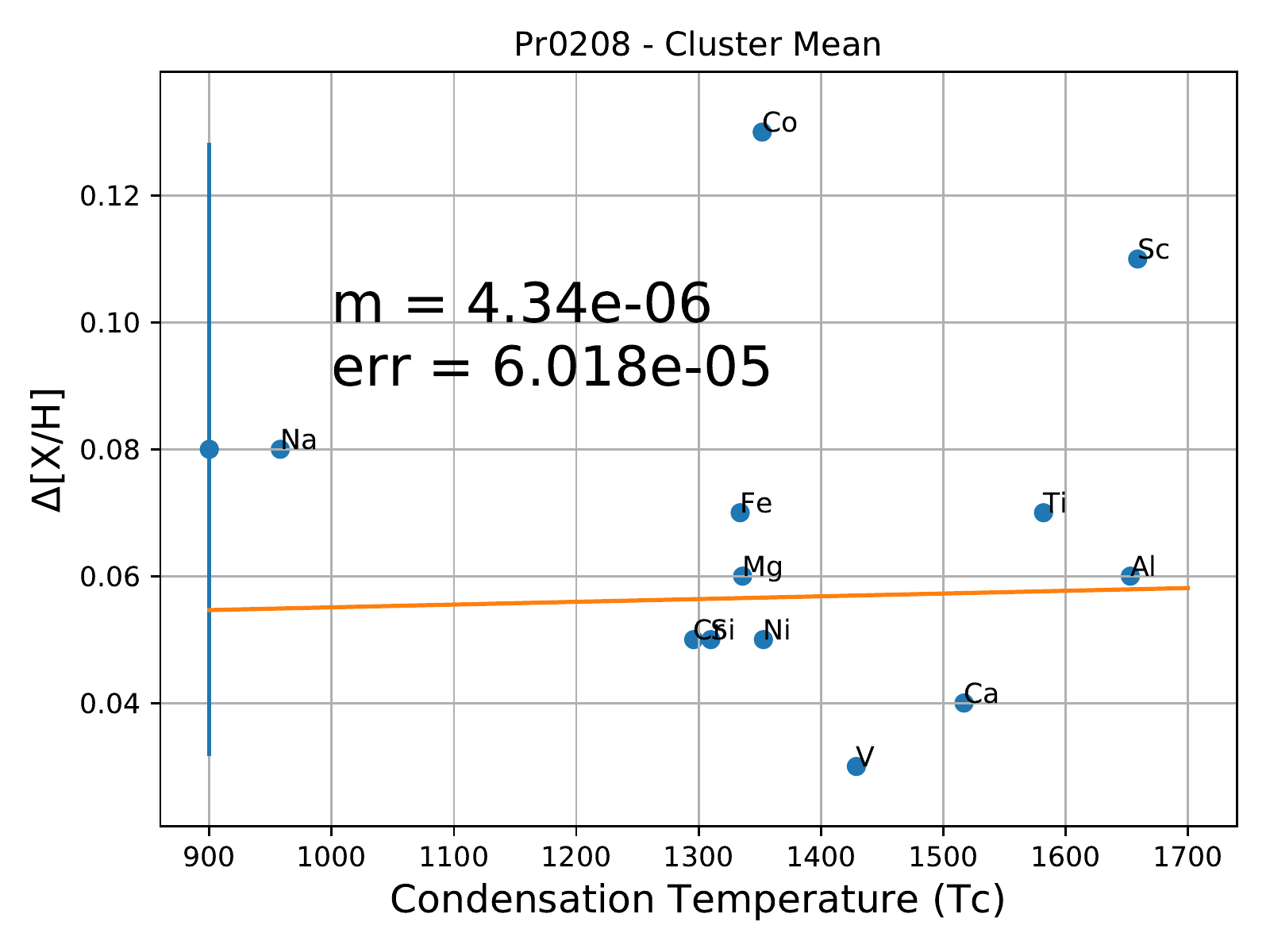}
\includegraphics[width = 3 in]{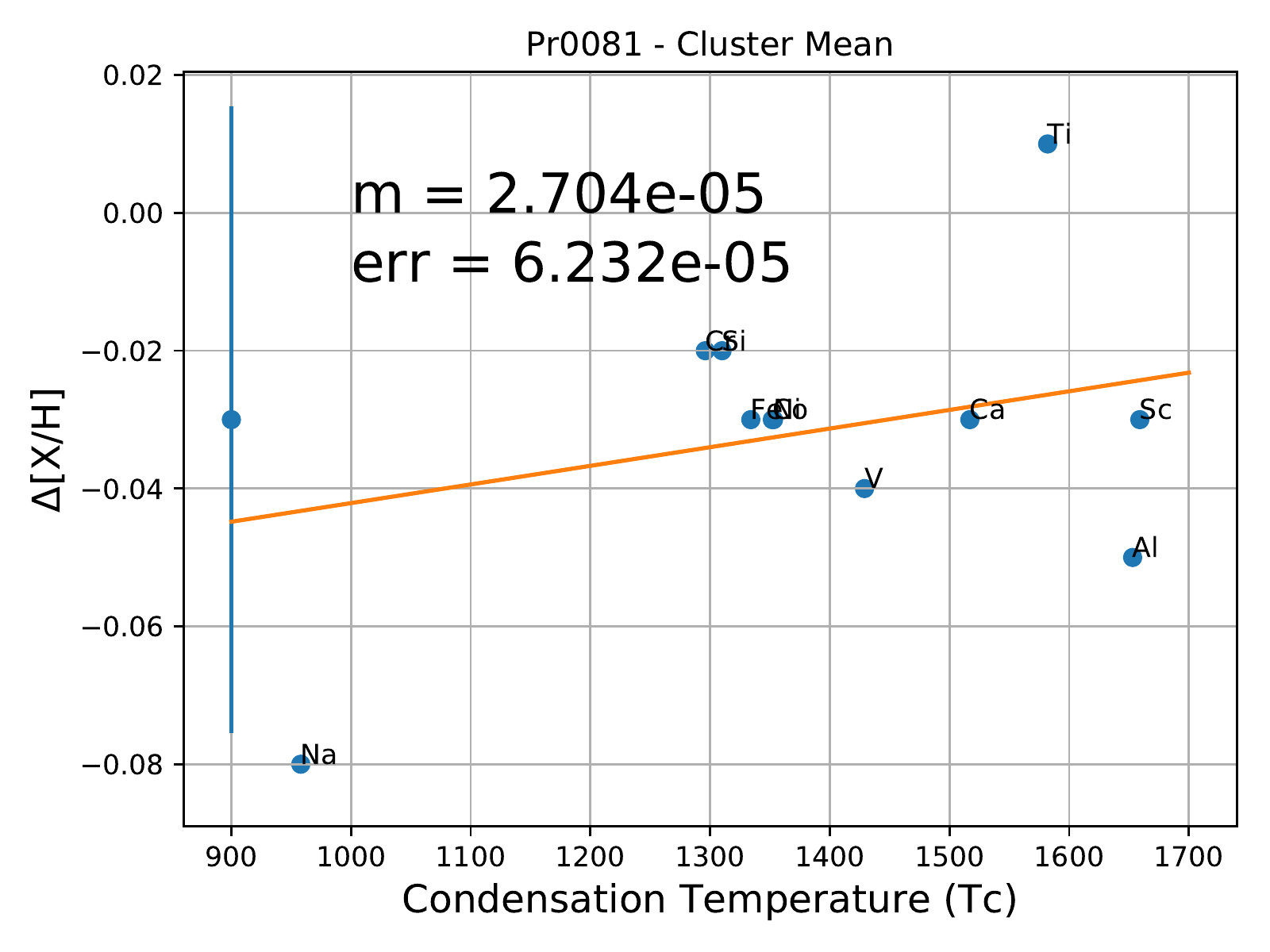}
\includegraphics[width = 3 in]{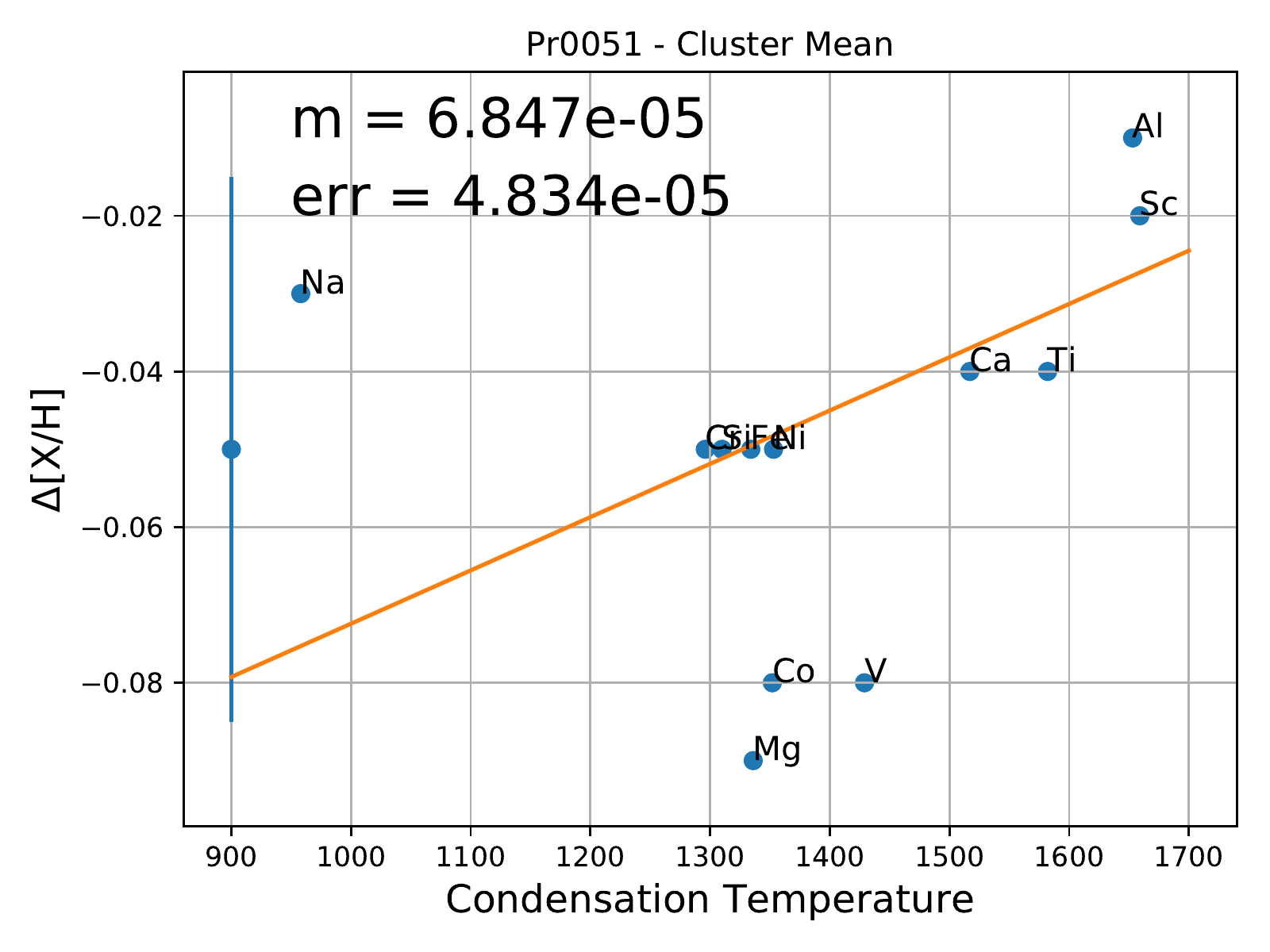}
\includegraphics[width = 3 in]{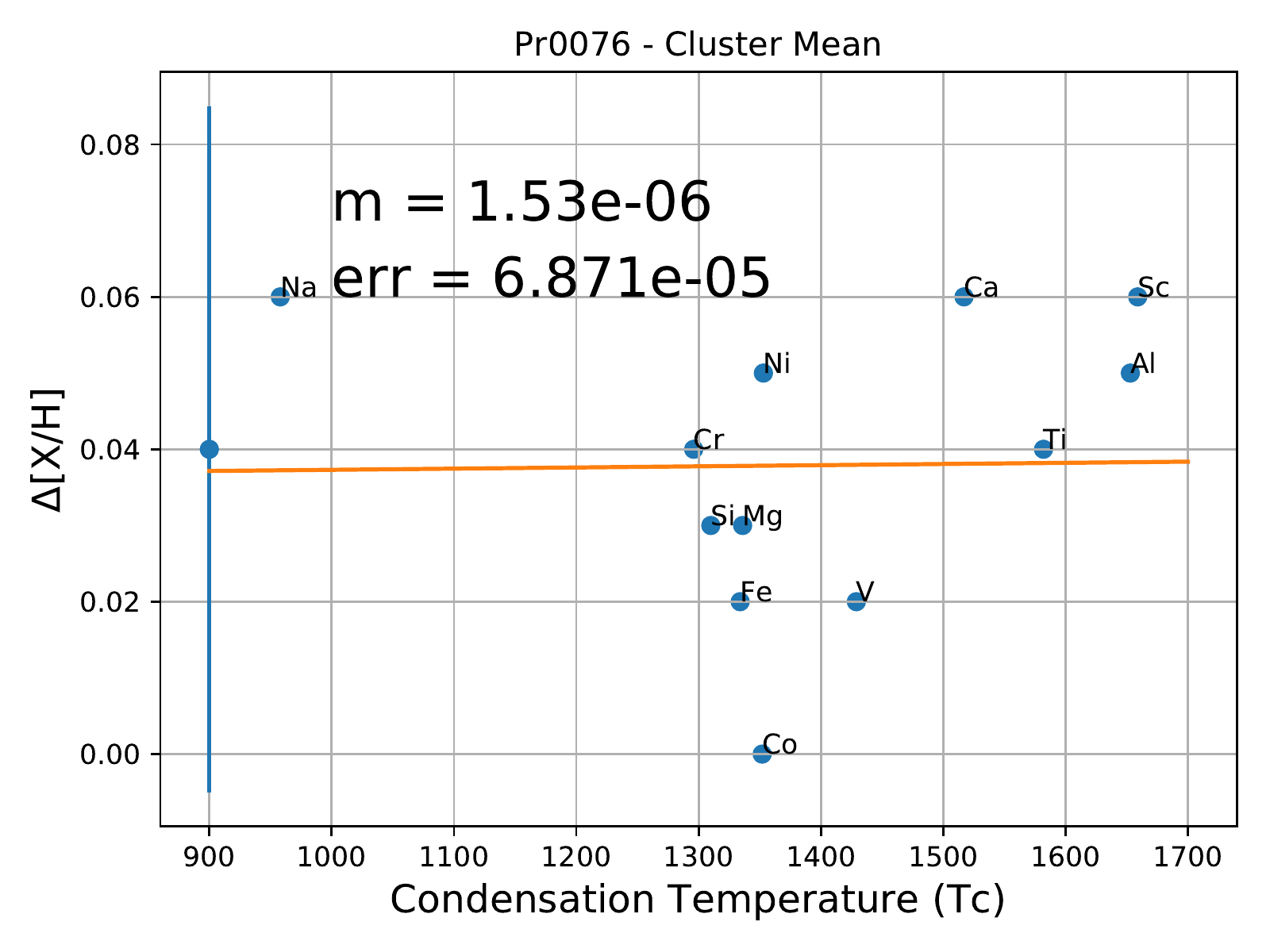}
\caption{\tc\ trend between each other star and cluster mean for refractory elements. The average error is shown on the left of each plot, which shows the $\pm1\sigma$ error with a dot to mark the center.}
\label{fig:others_mean_tc}
\end{figure}

Figure \ref{fig:mean_no_host_tc} shows the \tc\ trend for refractory elements of the cluster mean.
Here, we have excluded the planet host, Pr0201, to show that the mean cluster abundances alone present no trend with condensation temperature.
The cluster mean abundances show a slope of $-5.98 \times 10^{-6}$ $\pm$ $3.25 \times 10^{-5} \rm dex/K$ consistent with zero slope.
When comparing each other star with the cluster mean (excluding that star and the planet host in the mean), no star shows a statistically significant trend in refractory elements with condensation temperature (slopes shown in Fig \ref{fig:others_mean_tc}).

\begin{figure}[!ht]
\centering
\includegraphics[width=\linewidth,trim=0 10 0 25,clip]{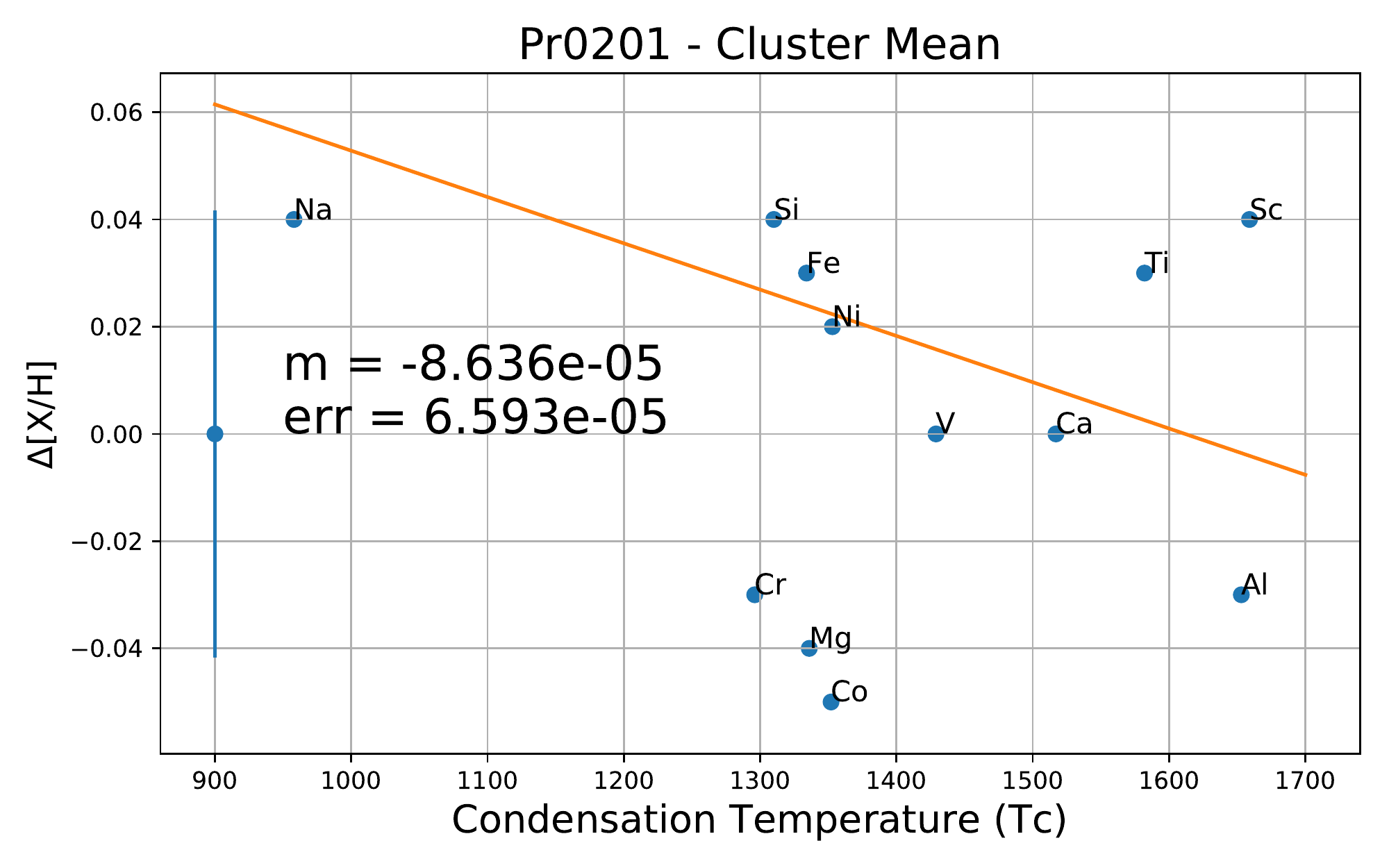}
\caption{\tc\ trend of the difference between planet host, Pr0201, and cluster mean for refractory elements. The average error is shown on the left side of the plot, which shows the $\pm1\sigma$ error with a dot to mark the center.}
\label{fig:host_mean_tc}
\end{figure}

We will take a closer look at Pr0201, because it is the only known planet-host in our sample. Figure \ref{fig:host_mean_tc} shows the \tc\ trend for refractory elements in the planet host Pr0201 relative to the cluster mean.
The best fit line to these elements gives a negative slope of $-8.64 \times 10^{-5}$ $\pm$ $6.59 \times 10^{-5} \rm dex/K$.
While not a statistically significant detection ($\rm 1.3\sigma$), if taken at face value, this trend could be explained by the sequestering of $\sim$1.62 \ME\ of material from the convection zone of Pr0201.

Our \tc\ slope results are relative to solar abundances ([X/H]), meaning a slope consistent with zero tells us the distribution of abundances with respect to \tc\ is similar to that of the sun.
Results for the cluster mean \tc\ slope are consistent with zero slope meaning the cluster abundance distribution may be similar to the sun.
This could indicate that planet-formation for sun-like stars in the cluster is common or that the cluster initially formed from a cloud with this distribution already in place.
The possible discovery of more planets within the cluster could give more weight to the prominent planet-forming case.
All six of our analyzed stars, including the known planet-host, show a slope consistent with zero when compared to the cluster mean.
If planet formation is in fact prevalent, the lack of detected planets in our sample could be due to the difficulty of finding smaller planets or these systems may have lost their planets entirely.

\subsection{Limits on Chemical Signatures of Planet Ingestion in Pr0201\label{s:tc_slopes}}

\begin{figure}[!ht]
\centering
\includegraphics[width = 4 in]{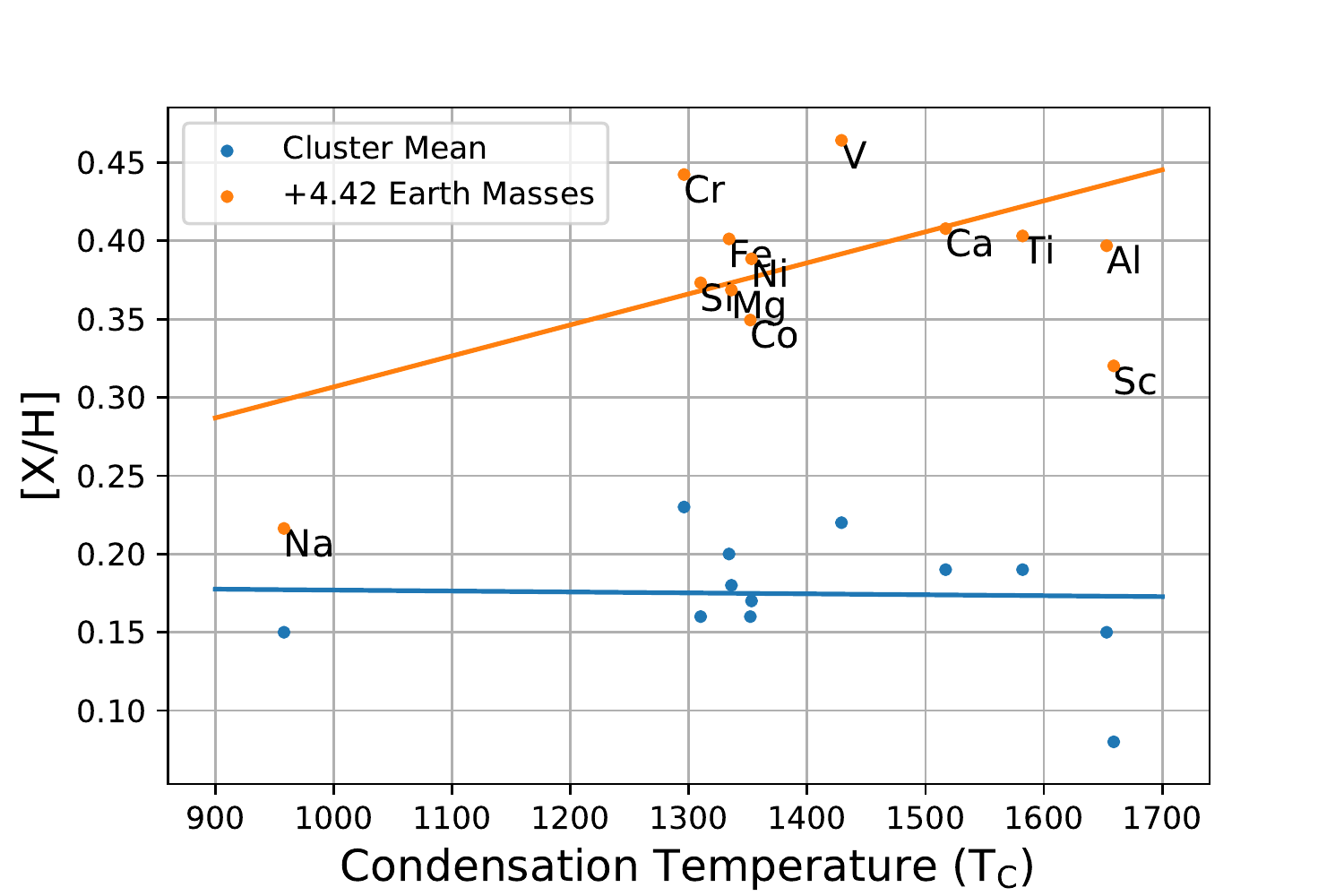}
\includegraphics[width = 4 in]{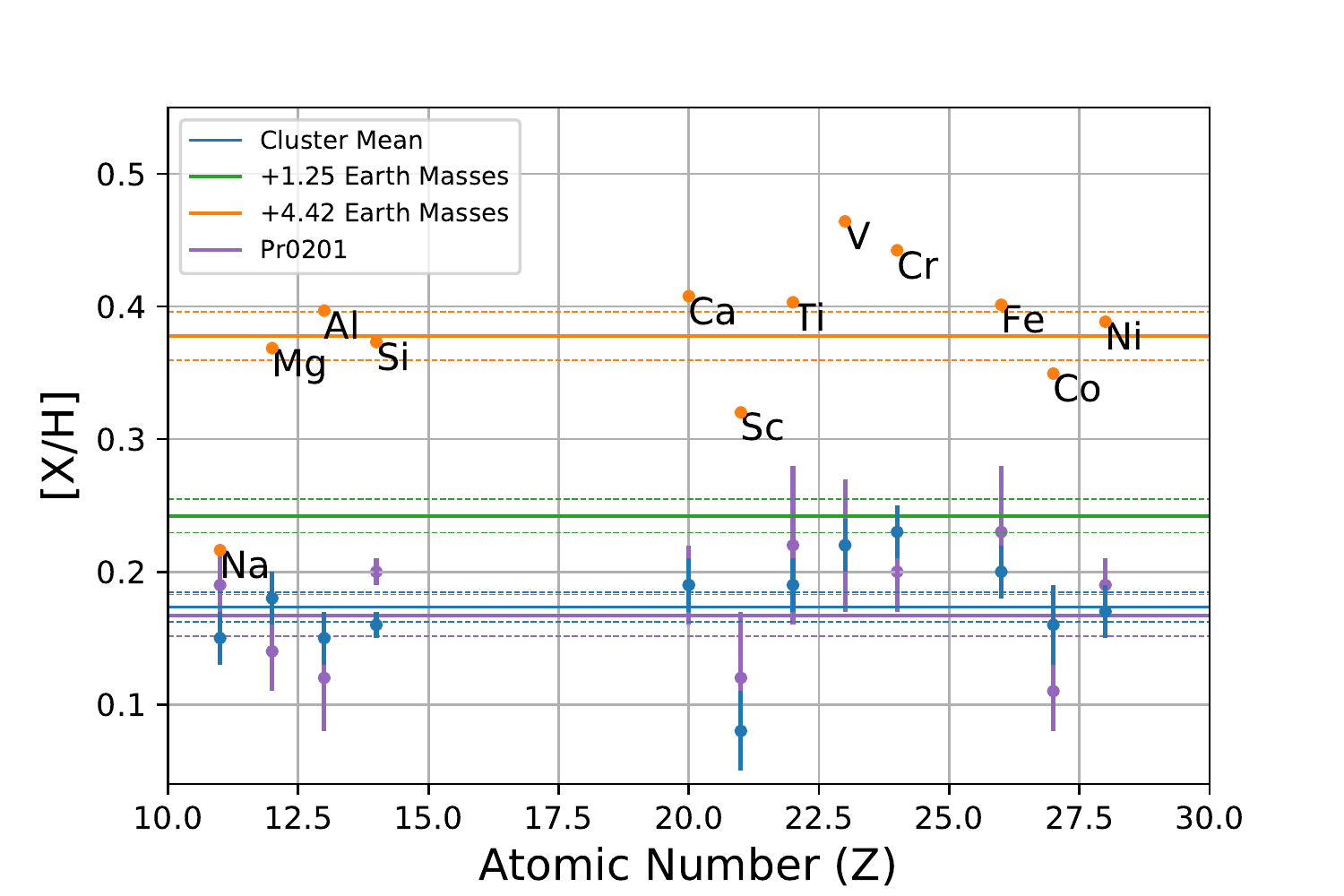}
\caption{Top panel: The refractory cluster mean abundances (excluding Pr0201) are shown in blue vs \tc\ along with simulated abundances (orange) of the addition of 4.42 \ME to the cluster mean, which would produce a statistically significant slope at the 3 sigma level. Bottom panel: Cluster mean abundances are shown in blue vs atomic number along with the average (solid line) and $\pm$ the error in the mean (dashed lines). Green and orange lines denote the addition of different amounts of \ME. Purple points show the abundances for Pr0201, the planet host.}
\label{fig:201_accretion}
\end{figure}

The Pr0201 system is host to a short period (4.4264 $\pm$ 0.0070 days) gas giant (Pr0201b) with a minimum mass of 0.54 $\pm$ 0.039 \MJ\ in a circular orbit \citep{2012ApJ...756L..33Q}.
With such a close-in giant planet and a \tc\ slope consistent with zero, we investigate the possibility that Pr0201 could have accreted or sequestered refractory-rich material during the planet formation process.
Using our planet engulfment model mentioned above, we can place limits on the amount of this material.
To do this, we assume Pr0201 formed with the same composition as the cluster mean, starting off with zero slope in the refractory elements.
We then add material with the same composition as the Earth until we produce a significant \tc\ slope ($\rm 3\sigma$ using the measured error in the \tc\ slope for Pr0201) in the refractory elements (shown in the top panel of Figure \ref{fig:201_accretion}).
In order to produce a statistically significant \tc\ slope, Pr0201 would have needed to accrete 4.42 \ME\ of material.
In the bottom panel of Figure \ref{fig:201_accretion}, we show how this would affect the individual refractory elements of the cluster mean, depicted by the orange points and lines, which can be compared to the abundances of Pr0201 in purple.
The solid lines show the mean abundance while the dashed lines show $\pm$ the error in the mean.
This accretion scenario would produce abundances that are noticeably enhanced compared Pr0201 and can be ruled out at the $\rm \sim16.5\sigma$ level.
Excluding Ti, V, and Fe, it is also the case that 1.25 \ME\ of material can be ruled out at the $\rm 5\sigma$ level, indicated by the green lines.
In general, Pr0201 does not seem significantly enhanced in refractory elements when compared to the cluster mean.
This analysis focuses on accretion but can also be applied to material sequestered.

%\section{Discussion\label{s:disc}}

\section{Conclusion\label{s:conc}}
In this work we have used new Keck/HIRES observations combined with KOA spectra of six G or F type stars, one of which hosts a 0.54 $M_{J}$ giant planet, in the nearby Praesepe Cluster to derive detailed solar-relative elemental abundances with a precision of $\sim$0.05 dex.
For each star we determined $\rm T_{\rm eff}$, $\log g$, [Fe/H], $\xi$ (microturbulence parameter), and abundances of up to 20 elements (Table \ref{tab:params}; Figure \ref{fig:abd_vs_z}).
We verify that our results are in good agreement with the current literature and determine a mean cluster metallicity of $+0.21 \pm 0.02$ dex.

We made use of a new custom-built python code for EW measurements called XSpect-EW.
The code automatically normalizes each order, wavelength shifts the orders to the rest frame, and fits a Gaussian profile to each line of interest while allowing the user to edit and rerun any of these processes with specified parameters when necessary. 
This makes measuring hundreds of lines much faster as many lines can be measured automatically with little user contribution and helps to remove some user error when manually performing these tasks.

We find no \tc\ trend in the mean cluster abundances (Figure \ref{fig:mean_no_host_tc}). 
Comparing each star's individual elemental abundances with the cluster mean abundances we see a negative \tc\ trend in the planet host, Pr0201, of $-8.64 \times 10^{-5}$ $\pm$ $6.59 \times 10^{-5} \rm dex/K$.
According to our planet engulfment model, the slope in Pr0201 corresponds to the sequestering of 1.62 \ME\ of terrestrial material from the convection zone of the star which could be an indication of terrestrial planet formation, although no terrestrial planets have been detected for this star.
We conclude that Pr0201 likely did not accrete a significant amount of Earth-like material.

As mentioned in the introduction, a natural dependence on cluster size and structure is prominent in determining the strength of the chaotic effects of the cluster environment on planet formation and survival.
Praesepe, being an open cluster, contains less mass than a globular cluster, only about $\sim$600 $\pm$ 19 $\rm M_{\odot}$ \citep{2002AJ....124.1570A}.
Less mass would dictate fewer O/B stars could initially form, lowering the effect of photoevaporation on disks, stellar winds and supernovae on gas expulsion, and impart a lower average velocity on the cluster members lowering the frequency of stellar encounters.
Even in high density environments, short period planets can survive throughout the lifetime of the cluster \citep{2019MNRAS.489.4311C}.
Cluster environmental effects on planet survivability are still not completely understood; however, the current picture painted by recent studies is one of planet formation being common and most of them surviving the evolution of the cluster \citep[albeit in smaller sized systems or as free-floating planets, most of which are expected to escape the cluster; e.g., ][]{2019A&A...624A.120V}.
Combined with the high metallicity of Praesepe, we have reason to believe that planets could be abundant in the cluster. 
Already, studies have found 13 planets in Praesepe and about 30 total planets in open clusters \citep{2019MNRAS.489.4311C}.
\citet{2020ApJ...897...60P} reported that our solar system likely formed in a high-mass extended or intermediate-mass compact association like NGC 6611 or Praesepe, due to roughly $10\%$ of solar-type stars experiencing flyby's after which a solar system analog would remain.
As RV surveys become more precise, we expect many more planets will be discovered in nearby open clusters.

\acknowledgements
The authors acknowledge support by {NSF AAG AST-1009810 and NSF PAARE AST-0849736}. This research has made use of the Keck Observatory Archive (KOA), which is operated by the W. M. Keck Observatory and the NASA Exoplanet Science Institute (NExScI), under contract with the National Aeronautics and Space Administration.
This work makes use of the Radial Velocity Simulator accessed through the Astronomy Education at the University of Nebraska-Lincoln Web Site (http://astro.unl.edu).
We thank the LSSTC Data Science Fellowship Program which has benefited this work.
George Vejar also thanks Karl Jaehnig for his help and support which benefited this work.

%
%Facilities: Keck~I (HIRES)
%
%
%%%%%%%%%%%%%%%%%
% REFERENCES
%%%%%%%%%%%%%%%%%
\newpage
%%The Bibliography

\end{document}